\documentclass[showpacs, amsmath, reprint, nofootinbib, showkeys, pra, aps, longbibliography]{revtex4-2} 
\usepackage{dcolumn}    
\usepackage{bm}         
\usepackage{hyperref}
\hypersetup{%
    colorlinks=true, 
    linkcolor=blue,
    urlcolor=blue,
    citecolor=blue, 
}

\usepackage{amssymb,lineno,amsfonts}
\usepackage{graphicx} 
\usepackage{bbm}
\usepackage{setspace}
\usepackage{float}
\usepackage{natbib}
\usepackage{array}
\usepackage{multirow}
\usepackage[normalem]{ulem}
\usepackage{booktabs}

\usepackage[utf8]{inputenc}
\usepackage{latexsym,amsfonts,amssymb,amsthm,amsmath}

\newcommand{\ket}[1]{\left\vert{#1}\right\rangle}
\newcommand{\bra}[1]{\left\langle{#1}\right\vert}
\newcommand{\outpr}[2]{\left\vert{#1}\right\rangle\left\langle{#2}\right\vert}
\newcommand{\inpr}[2]{\left\langle{#1}\middle\vert{#2}\right\rangle}

\newcommand{\tr}{\mathrm{Tr}}

\begin{document}

\title{Distributed Quantum Dense Coding Enhanced With Non-classical Routing}

\author{Pratham Hullamballi\textsuperscript{1, 2}}
\email{prathamprabhuhullamballi@gmail.com}

\author{Aparajita Bhattacharyya\textsuperscript{2}}
\email{aparajitabhattacharyya@hri.res.in}

\author{Ujjwal Sen\textsuperscript{2}}
\email{ujjwal@hri.res.in}

\affiliation{\textsuperscript{1}Department of Physics, Indian Institute of Science Education and Research, Pune 411008, India}

\affiliation{\textsuperscript{2}Harish-Chandra Research Institute, A CI of Homi Bhabha National Institute, Chhatnag Road, Jhunsi, Prayagraj 211 019, India}

\begin{abstract}
We propose a distributed quantum dense coding protocol that uses a control system to superpose two dense coding processes, allowing us to simultaneously and coherently encode and non-classically route the sender's single-qubit system to two receiver labs. We find that dense coding with coherently controlled encoding and routing performs better than the standard dense coding protocol in both global and one-way local decoding strategies employed by receiver labs in weak entanglement regimes and with noisy mixed states. We extend the protocol to incorporate noisy channels and show that, there exists a certain noise strength below which our protocol outperforms the noiseless standard dense coding protocol in presence of quantum dephasing noise. We extend our protocol to an arbitrary number of receivers and analyse its performance under a global decoding strategy. 
\end{abstract}

\maketitle

\section{Introduction}
The development of communication protocols has traditionally been rooted in classical information theory, where data is encoded and transmitted using classical bits. However, the advent of quantum mechanics has fundamentally altered the foundations of information science by providing us with the quantum extension of Shannon's classical information theory, arguably a more general framework for communication theory~\cite{wilde2013quantum}. By utilising non-classical resources like entanglement and superposition, communication protocols offer capabilities far beyond classical limits~\cite{De2011May}. 

Any communication scheme, quantum or classical, for sending classical information involves three major steps - (1) encoding of classical information into a physical system, (2) sending of the physical system through a physical channel, and (3) decoding of classical information from the received physical system. Communication protocols that involve encoding and decoding classical information through quantum states shared between distant parties are broadly referred to as quantum-assisted classical communication~\cite{Ziman2003Apr, Bose2000Feb}. Here, the physical system is a quantum system, and the physical channel is a quantum channel. In a seminal paper from Holevo~\cite{holevo1973bounds}, a hard bound on the amount of classical information extractable from a quantum system was derived. It proved that the amount of classical information that can be sent via $d$-dimensional quantum system is at most $\log_2{d}$ bits, therefore implying one cannot decode more than $n$ bits of information from $n$ qubits. The first instance of a quantum-assisted classical communication protocol giving an advantage over existing classical protocols was introduced by Bennett and Wiesner in 1992~\cite{Bennett1992Nov}, called quantum dense coding, where they used pre-shared entanglement between the sender and receiver as a resource to communicate 2 bits of information by transmitting only a single qubit, seemingly ``surpassing" the Holevo bound. The trick lies in pre-sharing a Bell state, thereby increasing the dimension of the quantum system and, consequently, the value of the Holevo bound. Further works have generalised dense coding protocol with mixed states of arbitrary dimension~\cite{Liu2002Jan, Bruss2004Nov}, multiple senders and up to 2 receivers~\cite{Bruss2004Nov, Bruss2006Jun}, and noisy channels~\cite{Das2015Nov, Shadman2010Jul}. To characterise the maximum potential of a dense coding protocol for a given state, the concept of dense coding capacity (DC capacity) was introduced. DC capacity proved to be helpful at characterising the role of entanglement and explicitly showed that bound-entangled states are not useful for dense coding~\cite{Bruss2004Nov}. Quantum dense coding has been experimentally realised in many quantum hardware setups, including atomic systems~\cite{Schaetz2004Jul}, photonic qubits~\cite{Mattle1996Jun, Barreiro2008Apr, Kong2017Sep, Williams2017Feb, Hu2018Jul}, optical modes~\cite{Li2002Jan, Jing2003Apr}, and nuclear magnetic resonance (NMR)~\cite{Fang2000Jan, Wei2004Mar}.

In recent years, there has been a growing interest in using a coherent quantum system~\cite{ Araujo2014Sep, Friis2014Mar} to do mainly two things - (1) put the causal ordering of any two quantum channels into a superposition, in an approach known as `quantum \texttt{SWITCH}'~\cite{Chiribella2013Aug, Ebler2018Mar}, rendering the causal order of the channel `indefinite'~\cite{Goswami2021Apr} and (2) to coherent control quantum channels themselves, usually referred to as channel multiplexing or a `quantum half-\texttt{SWITCH}'~\cite{Abbott2020Sep}. The latter approach was first introduced in error filtration~\cite{Gisin2005Jul}. These approaches show classical and quantum capacity enhancement when used in communication protocols~\cite{Chiribella2019May, Abbott2020Sep, Ebler2018Mar}. It was discovered that one could transmit non-zero information using a quantum state passing through either a quantum \texttt{SWITCH} or quantum half-\texttt{SWITCH} of two completely depolarising channels~\cite{Abbott2020Sep, Ebler2018Mar}. In a complete surprise, perfect quantum communication with two completely depolarising channels using quantum $\texttt{SWITCH}$ was proven to be possible~\cite{Chiribella2021Mar}. Further resource-theoretic studies on the superposition of quantum trajectories and their role in communication protocols have also been done~\cite{Kristjansson2020Jul, Vanrietvelde2021Jun}. Experimental attempts using interferometric setups have been made to explore communication advantages using superposition of trajectories~\cite{Procopio2015Aug, Rubino2017Mar, Rubino2022Jan, Goswami2018Aug, Goswami2018Jul, Wei2019Mar, Guo2020Jan, Taddei2021Feb, Rubino2021Jan}. 

Inspired by the possibility of using a coherent quantum system as a control to superpose two situations, we will use a coherent quantum system to superpose two dense coding scenarios, allowing coherently controlled encoding and ``non-classical" routing of the sender's system to two receiver labs. 
We investigate the DC capacity when the two receiver labs are together and find a lower bound to the ``local'' DC capacity when they are distant but can do one-way classical communication. We also extend it to the distributed case involving an arbitrary number of receiver labs. We find that our protocol performs better than the standard dense coding protocol in both global and one-way LOCC decoding scenarios in weak entanglement regimes and with mixed states. We further extend our protocol to account for the presence of noisy channels. In particular, we demonstrate that, under quantum dephasing noise, there exists a threshold noise strength below which our protocol surpasses the performance of the standard dense coding protocol, even when the latter is implemented in an ideal  (noiseless) setting.

We present the work in the following way. In section~\ref{sec:standard_dense_coding}, we describe the standard dense coding protocol. In section~\ref{sec:unitary_multiplexing}, we describe the approach of unitary multiplexing using a control system. In section~\ref{sec:ourprotocol}, we present our variant of the dense coding protocol using unitary multiplexing for coherent encoding and non-classical routing and explain our results under global and one-way local decoding scenarios. The detailed analysis of the noisy distributed dense coding scenario is presented in Sec.~\ref{sec:noise}. In section~\ref{sec:distributed}, we consider multiplexing in the case of multiple receivers. Finally, we summarise in section~\ref{sec:ending}.

\section{Standard Dense Coding Protocol} \label{sec:standard_dense_coding}
The main idea of dense coding is to utilise a previously shared entanglement between a sender and a receiver, aiming to transmit more information than would be feasible without using entanglement as a resource. The standard dense coding protocol involving a single sender and single receiver is discussed below.

\subsection{Single-sender and single-receiver case} \label{subsec:single_sender_single_receiver}
Consider a single sender, Alice ($A$), and a single receiver, Bob ($B$),
sharing a quantum state $\rho_{AB} \in \mathbb{C}^{d_A} \otimes \mathbb{C}^{d_B}$, 
where $\mathbb{C}^{d_{A(B)}}$ denotes the complex Hilbert space of the subsystem, $A(B)$, having dimension $d_{A(B)}$. Let $X$ be a random variable whose realisations $x$ are \textit{letters} in an \textit{alphabet} $\mathcal{X}$. A general dense coding protocol can be described in three steps~\cite{Bruss2004Nov}: 
\begin{enumerate}
    \item \textbf{Encoding}: Alice encodes an information as a random variable $X$ on her part of $\rho_{AB}$. The encoding process involves the application of a local unitary operator, $U_A^x$, with probability $p_X(x)$ on Alice's part of the subsystem. As a result, the state $\rho_{AB}$ gets converted to an encoded ensemble $\overline{\mathcal{E}} = \{p_X(x), \rho_{AB}^x \}$, where $\rho_{AB}^x = (U_A^x \otimes \mathcal{I}_{d_B})\rho_{AB}({U_A^x}^{\dagger} \otimes \mathcal{I}_{d_B})$ and $x \in \mathcal{X}$ are realisations of the random variable $X$. 
    \item \textbf{Sending}: Alice deterministically sends \textit{her part} of the ensemble state to Bob through a quantum channel. For our purpose, we consider it to be a noiseless quantum channel. 
    \item \textbf{Decoding}: Since Alice sends her part of the ensemble to Bob, the ensemble, $\overline{\mathcal{E}} = \{p_X(x), \rho_{AB}^x \}$, is now with Bob. He tries to extract maximal information about the random variable $X$ from the ensemble by performing suitable measurements,  $\{\Lambda_y^{B}\}_y$ where $y$ are realisations of the random variable $Y$, and the maximisation of mutual information, $I(X;Y)$, is performed over all positive operator valued measurements.
\end{enumerate}
The maximal mutual information - maximum over measurements performed by the receiver - also referred to as the accessible information of the ensemble $I_{acc}(\overline{\mathcal{E}}) = \max_{\{\Lambda_y^{B}\}_y} I(X;Y)$, is upper bounded by the Holevo quantity $\chi(\overline{\mathcal{E}})$ given by
\begin{equation}
    I_{acc}(\overline{\mathcal{E}}) \le \chi(\overline{\mathcal{E}}) = S(\rho_{\overline{\mathcal{E}}}) - \sum_{x \in \mathcal{X}} p_X(x) S(\rho_{AB}^x).
\end{equation}
where $\rho_{\overline{\mathcal{E}}} = \sum_x p_X(x) \rho_{AB}^x$ is the density matrix representing the encoded ensemble. Optimising the Holevo quantity over all possible unitary encodings gives us the dense coding capacity $\chi_{DC}$~\cite{Bruss2004Nov}
\begin{equation}
    \chi_{DC} = \log(d_A) + S(\rho_B) - S(\rho_{AB}).
    \label{eq:globalDC}
\end{equation}
The optimal unitary encodings and their corresponding probabilities, $\{p_X(x), U_A^x\}$, which maximise the Holevo quantity is given by $\{\frac{1}{d_A^2}, M_k\}$. Here, the set, $\{M_k\}_{k=0}^{d_A^2-1}$, is a complete set of mutually orthogonal unitary operators that satisfy the relation $\frac{1}{d_A^2}\sum_k M_k^{\dagger} Q M_k = \tr(Q)$ for any operator $Q$. An example that satisfies these conditions is given by the group of shift-and-multiply operators~\cite{Hiroshima2001Aug},
\[ M_{(p, q)} \ket{j} = \exp\left\{i\frac{2 \pi}{d}pj \right\} \ket{j + q(\text{mod $d_A$})}, \]
where $k\equiv(p, q)$ and $p, q, j \in \{0, 1, \cdots, d_A-1\}$. It is proven that the Holevo bound is asymptotically achievable, in a sense that if the sender is able to send long strings of the encoded state, then there exists a particular encoding and a decoding scheme that asymptotically saturates the bound~\cite{Bruss2006Jun}. Therefore, $\overline{\chi}_{DC}$ can be asymptotically saturated.

\subsection{Single-sender and two-receivers case} \label{subsec:single_sender_two_receiver}
In order to compare the standard dense coding protocol with the scenario that we will soon consider in the following sections, it is convenient to inspect the usual protocol involving a single sender, Alice, and two receivers, Bob$_1$ and Bob$_2$. For a global decoding scenario, the two-receiver case can effectively be considered as a single receiver $B \equiv B_1B_2$. Let us suppose that in the two-receiver scenario, Alice, Bob$_1$ and Bob$_2$ share a generalised Greenberger-Horne-Zeilinger (GHZ) state $\ket{gGHZ}_{AB_1B_2} = \cos(\theta/2)\ket{000} + e^{i\phi} \sin(\theta/2)\ket{111}$, where $\theta \in [0, \pi]$ and $\phi \in [0, 2\pi]$. Since $d_A = 2$, the optimal unitary 
operators used for encoding are given by $\{\mathcal{I}, \sigma_z, \sigma_x, \sigma_x\sigma_z\}$, each of which are applied with equal probabilities. Following this, Alice deterministically sends her part of the qubit to party $B$. As a result, an encoded ensemble $\overline{\mathcal{E}}$ is created at the receivers' end. The information that the two receivers are able to extract if they have access to global measurements is referred to as the globally accessible information  $I_{acc}^{glo}$~\cite{Suzuki2007Sep, Preskill2016Apr}. There exists a useful upper bound to globally accessible information given by the Holevo quantity $\chi(\overline{\mathcal{E}})$. For pure states, the Holevo quantity simplifies to $\chi(\overline{\mathcal{E}}) = S(\rho_{\overline{\mathcal{E}}})$ where $S(\rho_{\overline{\mathcal{E}}})$ denotes the von Neumann entropy of the density matrix belonging to the encoded ensemble $\overline{\mathcal{E}}$. It can be shown that the dense coding capacity in this scenario is
\begin{multline}
    \chi_{SDC} \equiv \chi(\overline{\mathcal{E}}) = 1-2\bigg\{ \cos^2\left(\frac{\theta}{2}\right) \log\left[ \cos\left(\frac{\theta}{2}\right)\right] + \\ \sin^2\left(\frac{\theta}{2}\right) \log\left[ \sin\left(\frac{\theta}{2}\right)\right]  \bigg\}.
    \label{eq:normalDC}
\end{multline}
where SDC refers to Standard Dense Coding.

The above result holds even when we consider a situation with a single sender, Alice, and multiple receivers, Bob$_1$ $\cdots$ Bob$_{M-1}$, sharing a $M$-dimensional GHZ state because, here too, we can take $B\equiv B_1\cdots B_{M-1}$. 

The dense coding capacity in this scenario is independent of the phase, is symmetric about $\theta = \frac{\pi}{2}$ and reaches the maximum of $\chi_{max} = 2$ at $\theta = \frac{\pi}{2}$ for a generalised GHZ state. Unless Alice possesses at least two qubits of the GHZ state, there are no local operations Alice can perform that can reach all eight mutually orthogonal states and go past two bits of dense coding capacity~\cite{Cereceda2001May}.

\section{Unitary Multiplexing}~\label{sec:unitary_multiplexing}
Before we get into the concept of non-classical routing in a dense coding protocol, it will be useful to introduce the method of multiplexing unitaries with the aid of a coherent quantum system. This will help us in the eventual compact, mathematical formulation of our protocol. 

Consider two given unitary operators, $U$ and $V$, a target system $\rho_t$ and a control system $\rho_c$. Mathematically, one can construct a controlled unitary operator of the form, $W = U \otimes \ket{0}\bra{0}_c + V \otimes \ket{1}\bra{1}_c$, which acts on the composite state of the system and control~\cite{Abbott2020Sep}. If we set the control to be $\rho_c = \ket{+}\bra{+}$, then the joint state of the system and control obtained after unitary evolution is
\begin{eqnarray}
     \mathcal{N}(U, V)(\rho_t \otimes \rho_c) &=& W(\rho_t \otimes \rho_c)W^{\dagger} \nonumber \\
     &=& \frac{1}{2} \bigg(  U\rho_t U^{\dagger} \otimes \outpr{0}{0}_c +   U \rho_t V^{\dagger} \otimes \outpr{0}{1}_c \nonumber \\&+&   V \rho_t U^{\dagger} \otimes \outpr{1}{0}_c +  V \rho_t V^{\dagger} \otimes \outpr{1}{1}_c \bigg). \nonumber
\end{eqnarray}  
If we consider pure input, i.e. $\rho_t = \outpr{\psi}{\psi}$, we observe that the above equation simplifies to
\begin{equation}
\label{coh_control}
    \mathcal{N}(U, V)(\rho_t \otimes \rho_c) = \ket{\Psi}\bra{\Psi}
\end{equation}
where $\ket{\Psi} = \frac{1}{\sqrt{2}} \big( U\ket{\psi} \otimes \ket{0} + V\ket{\psi} \otimes \ket{1} \big)$. Now, a measurement is performed on the control system in the Hadamard basis, $\{\outpr{+}{+}, \outpr{-}{-} \}$, and the outcome corresponding to $\outpr{+}{+}$ is post-selected. This causes the target system to collapse into the unnormalised state $\ket{\phi}$, given by
\begin{equation}
    \ket{\phi} = \frac{1}{2} (U+V) \ket{\psi}.
\end{equation}
Using this strategy, we can thus multiplex unitaries $U$ and $V$ using a control quantum system. We will employ such multiplexing to model our dense coding protocol provided in the succeeding section.

\section{Dense Coding With Non-Classical Routing} \label{sec:ourprotocol}
Consider a single sender, Alice (A), and two receivers, Bob$_1$ ($B_1$) and Bob$_2$ ($B_2$). They share an arbitrary tripartite pure state $\ket{\psi}_{AB_1B_2}$. Let Bob$_1$ and Bob$_2$ have access to auxiliary systems $\ket{0}_{C_{B_1}}$ and $\ket{0}_{C_{B_2}}$, which they can use whenever required. The two receivers, Bob$_1$ and Bob$_2$, reside in laboratories $L_1$ and $L_2$, respectively. Depending on to whom Alice wants to send her quantum system, the following two situations arise:
 
\begin{itemize}
    \item \textbf{Process 1}: Alice applies local unitaries $U^x$ with probability $p_X(x)$ to her part of the tripartite state and sends her quantum system to lab $L_1$ via a noiseless quantum channel. Once sent, we formally define the labs as \textit{ordered} quantum systems $L_1 \equiv (B_1, A)$ and $L_2 \equiv (B_2, C_{B_2})$. The ensemble possessed by receivers is
    \begin{eqnarray}
       \;\;\;\;\;\;\;&&\left\{p_X(x), U^x_2 \ket{\psi}_{L_1L_2} \right\}\nonumber \\
       &&\equiv \left\{p_X(x), (\mathcal{I}_{B_1} \otimes U^x \otimes \mathcal{I}_{B_2} \otimes  \mathcal{I}_{C_{B_2}}) \ket{\psi}_{B_1A:B_2C_{B_2}} \right\}. \nonumber
    \end{eqnarray}
    This configuration is associated with the control state $\rho_c = \outpr{0}{0}$.

    \item \textbf{Process 2}: Alice applies local unitaries $V^x$ with probability $p_X(x)$ to her share of the tripartite state and sends it to lab $L_2$ via a noiseless quantum channel. Once sent, like earlier, we define the two labs as \textit{ordered} quantum systems $L_1 \equiv (B_1, C_{B_1})$ and $L_2 \equiv (B_2, A)$. The ensemble possessed by the two receivers, in this case, is
    \begin{eqnarray}
    \;\;\;\;\;\;\;&& \left\{p_X(x), V^x_4 \ket{\psi}_{L_1L_2} \right\}\nonumber \\
       &&\equiv \left\{p_X(x), (\mathcal{I}_{B_1} \otimes \mathcal{I}_{C_{B_1}} \otimes \mathcal{I}_{B_2} \otimes V^x) \ket{\psi}_{B_1C_{B_1}:B_2A} \right\}. \nonumber
    \end{eqnarray}
    This configuration is associated with the control state $\rho_c = \outpr{1}{1}$.
    
\end{itemize}

\begin{figure*}[t]  
    \centering
    \includegraphics[width=\textwidth]{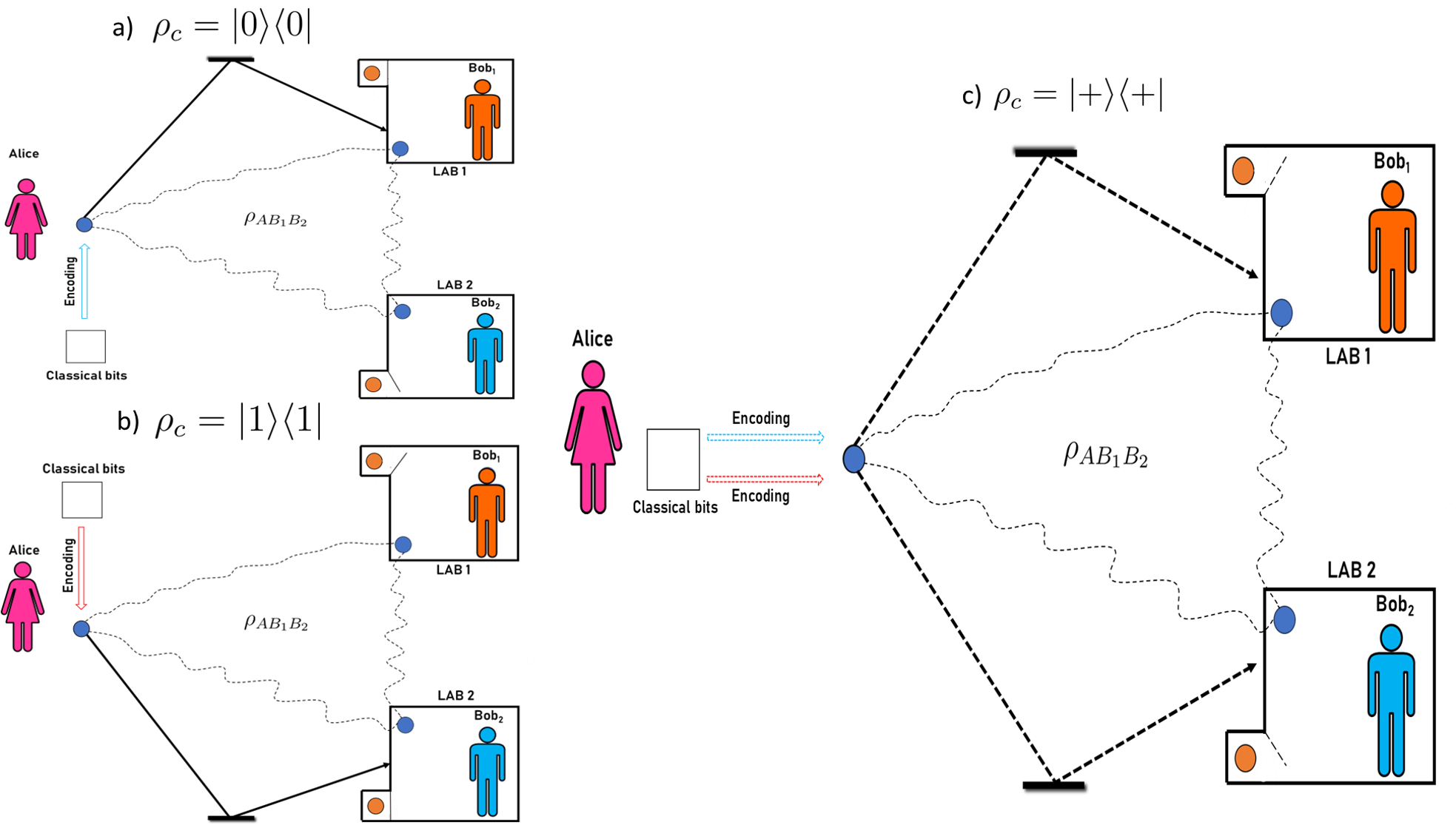}
    \caption{\textbf{a)} shows Process 1, where Alice encodes and routes her quantum system to lab $1$, and the orange ball in lab $2$ represents a fixed auxiliary state, $\ket{0}$. This is associated with the state $\rho_c = \outpr{0}{0}$. Similarly, \textbf{b)} shows Process 2, where Alice encodes and routes her system to lab $2$ with the auxiliary state, $\ket{0}$, in lab $1$. This is associated with the state $\rho_c = \outpr{1}{1}$. In \textbf{c)}, by setting $\rho_c = \outpr{+}{+}$, a superposition of Process 1 and 2 is created where the encoding and routing are coherently controlled. Here, it is important to note that the measurement outcome of the control quantum system needs to be sent via a noiseless classical channel with limited capacity to the receivers.}
    \label{fig:dc_ncr}
\end{figure*}

We aim to create a superposition of these two processes by setting the control state to be $\rho_c = \outpr{+}{+}$, in which case the sending itself becomes neither deterministic nor probabilistic but completely non-classical. Our idea is to employ another quantum system, referred to as the ``control", which coherently controls the above two processes. We will proceed with the assumption that the superposition of the above two processes is possible. Similar to the formalism provided in the preceding section, processes $1$ and $2$ are multiplexed with the help of a control system in the following way
\begin{eqnarray}
    \frac{1}{\sqrt{2}} \left( U^x_2 \ket{\psi}_{L_1L_2} \otimes \ket{0}_C + V^x_4 \ket{\psi}_{L_1L_2} \otimes \ket{1}_C \right),
\end{eqnarray}
where $C$ denotes the control qubit. This is followed by a measurement being performed on the control qubit in the Hadamard basis. Now, if the state collapses to $\ket{+}$ after the measurement, then the state possessed at the receivers' end is given by
\begin{equation}
    \ket{\Psi_x} =  \frac{1}{\sqrt{\inpr{\Psi_x}{\Psi_x}}} \left( U^x_2 \ket{\psi}_{L_1L_2} + V^x_4 \ket{\psi}_{L_1L_2}  \right).
    \label{eq:superposition}
\end{equation}
The state $\ket{\psi}_{L_1L_2}$ next to $V^x_4$ is the same as the one next to $U^x_2$ but with the 2nd and 4th indices swapped. Therefore
\begin{equation}
\label{ref00}
     \ket{\Psi_x} = \frac{1}{\sqrt{\inpr{\Psi_x}{\Psi_x}}} \left( U_2^x + V_4^x \cdot \texttt{SWAP}_{2,4} \right) \ket{\psi}_{L_1L_2}
\end{equation}
where $\texttt{SWAP}_{2,4}$ is a \texttt{SWAP} unitary acting on indices 2 and 4 occurring due to the routing of Alice's state. The $\texttt{SWAP}$ unitary arises naturally due to Alice choosing to route her system to either lab $L_1$ or $L_2$. 
Notice that processes $1$ and $2$ are individually identical (taking $U = V$), differing only by a trivial reordering; however, this ordering acquires non-trivial significance when a superposition of the two processes is considered.
The whole protocol is pictorially depicted in Fig.~\ref{fig:dc_ncr}. 
Figure~1 illustrates the three scenarios underlying our protocol. Throughout the figure, the blue ball represents Alice's encoded quantum system, while the orange ball denotes the auxiliary state $\ket{0}$.
In Fig.~1(a), Alice encodes the classical message into a quantum system and routes it to Lab~1 (Bob$_1$), while Lab~2 (Bob$_2$) receives the auxiliary state. This process corresponds to the control state $\rho_c=\ket{0}\bra{0}$.
In Fig.~1(b), the encoded quantum system is instead routed to Lab~2, while Lab~1 receives the auxiliary state. This process corresponds to the control state $\rho_c=\ket{1}\bra{1}$.
Finally, in Fig.~1(c), the control is prepared in the superposition state $\rho_c=\ket{+}\bra{+}$, resulting in a coherent superposition of the two routing configurations shown in Figs.~1(a) and 1(b). After the control system is measured, its outcome is communicated to the receivers through a noiseless classical channel of limited capacity.


In our protocol, the two receivers always possess systems of equal dimension to ensure that the superposition of the routing processes is well-defined within a common Hilbert space. In each individual process (before the superposition is created), if Alice sends her qubit to laboratory $L_1$ (Bob$_1$), then laboratory $L_2$ (Bob$_2$) does not remain empty; instead, it contains a fixed auxiliary qubit prepared in a predetermined reference state. Conversely, if Alice sends her qubit to $L_2$, then $L_1$ holds the corresponding auxiliary qubit in that reference state. This ensures that, irrespective of the routing choice, both laboratories are described by Hilbert spaces of identical dimension. Consequently, when we coherently superpose the two routing processes, the superposition is well-defined within a fixed global Hilbert space. No physical subsystem is removed or destroyed at any stage.\\

We further emphasize that this feature is an inherent consequence of quantum superposition, which underlies the quantumness of the protocol. The two routing possibilities are not realized exclusively, as in a classical scenario, but are instead coherently controlled and exist simultaneously at the level of amplitudes. In particular, it is the inherent nature of superposition that both routing paths are coherently controlled together, rather than one qubit being destroyed by the other. Thus, it is not that one qubit is destroyed; rather, both routing paths contribute coherently to a single, well-defined quantum state. The auxiliary systems ensure dimensional consistency, and the superposition enables interference between the alternatives, which is precisely what gives rise to the observed advantage. \\

\renewcommand{\arraystretch}{1.4}
\begin{table*}[t]
\centering
\begin{tabular}{|c|c|c|}
\hline
\textbf{Initial Shared State} & \textbf{Decoding Strategy} & \textbf{Observed Advantage} \\
\hline
\multirow{3}{*}{$\ket{gGHZ}$} 
  & Global (noiseless) & $\Delta>0$ for $\theta\in(0, 0.75]\cup [2.4, \pi)$ \\
\cline{2-3}
  & Global (dephasing noise) & $\Delta_{\text{noiseless}}^{\text{noisy}}, \Delta_{\text{noisy}}^{\text{noisy}}>0$ for $p\le0.03$
  \\
\cline{2-3}
  & $\texttt{LOCC}_1$ (only PVM measurements) & $\Delta>0.32$ for $\theta\in\{0.0147, \pi-0.0147\}$ \\
\hline
$\frac{1}{8}(\mathcal{I}_{A} \otimes \mathcal{I}_{B_1} \otimes \mathcal{I}_{B_2})$ 
  & Global (noiseless) & $\Delta \approx 0.2539$ bits \\
\hline
$\sum_{i=1}^{2} p_i \rho_A^i \otimes \rho_{B_1}^i \otimes \rho_{B_2}^i$ 
  & Global (noiseless) & $\Delta \approx \log(5/4)$ \\
\hline
\end{tabular}
\caption{Summary of the scenarios analyzed and the corresponding parameter regimes in which our non-classically routed (NCR) dense coding protocol outperforms the standard dense coding (SDC) protocol. The advantage is quantified by $\Delta = \chi_{NCR}^{glo}-(\chi_{SDC}+1)>0$. $\Delta_{\text{noiseless}}^{\text{noisy}}$ refers to the advantage of noisy NCR protocol over the same noiseless and noisy SDC protocols, respectively. The parameter $p$ is the dephasing parameter of quantum dephasing noise.}
\label{tab:summary}
\end{table*}
\renewcommand{\arraystretch}{1.0}

So far, we have seen that Alice encodes a pair of unitaries, $\{U^x, V^x\}$, via multiplexing and non-classically routes them to laboratories $L_1$ and $L_2$.  
This results in the state shared between $L_1$ and $L_2$ to be $\ket{\Psi_x}$. In the process, the encoded ensemble possessed by laboratories $L_1$ and $L_2$ is given by $\mathcal{E} \equiv \left\{  p_X(x), \rho_x = \outpr{\Psi_x}{\Psi_x}  \right\}$. The multiplexing action to get $\ket{\Psi_x}$ can be effectively modelled from a unitary operation on the joint state of the system and control, followed by performing a measurement on the control system in the Hadamard basis. The effective unitary operation is given by  $\mathcal{N}(U_2^x, V_4^x)(\rho \otimes \rho_c) = W_x(\rho \otimes \rho_c)W_x^{\dagger}$, where the global unitary is of the form
\begin{multline}
    W_x = (\mathcal{I}_{B_1} \otimes U^x \otimes \mathcal{I}_{B_2} \otimes \mathcal{I}_{C_{B_2}}) \otimes \outpr{0}{0} + \\ (\mathcal{I}_{B_1} \otimes \mathcal{I}_{C_{B_1}} \otimes \mathcal{I}_{B_2} \otimes V^x) \cdot \texttt{SWAP}_{2,4} \otimes \outpr{1}{1}.
\end{multline}

The map $W_x$ should be best thought of as an effective mathematical description, a ``supermap”, that captures the joint evolution of the target and the control system. Since the structure of the unitary, $W_x$, is known to us, it defines our protocol. Then, we can extend our protocol from pure states to an arbitrary state $\rho_{B_1AB_2}$, including mixed ones, where the ensemble possessed by the two labs is $\mathcal{E} = \{p_X(x), \rho_x\}$, such that
\begin{equation}
\label{ref:unitary}
    \rho_x = \tr_c\left[ \frac{(\mathcal{I}_t \otimes \Pi_+)(W_x(\rho_t \otimes \rho_c)W_x^{\dagger})(\mathcal{I}_t \otimes \Pi_+)}{\tr\left( (\mathcal{I}_t \otimes \Pi_+)(W_x(\rho_t \otimes \rho_c)W_x^{\dagger}) \right)}  \right]
\end{equation}
where $\rho_t = \rho_{B_1AB_2} \otimes \outpr{0}{0}$ and $\Pi_+ = \outpr{+}{+}$. The control is traced out because the receivers do not have access to it. It is important that the outcome of the measurement performed on the control qubit is classically communicated to the receivers. The transmission of this one bit of information happens through a noiseless classical channel. We will assume this noiseless classical channel has limited capacity, i.e., we can send one bit of information through it per quantum channel usage. We need to take this into account while comparing our protocol with the standard dense coding protocol. 

Given the ensemble $\mathcal{E} = \left\{ p_X(x), \rho_x \right\}$ with laboratories $L_1$ and $L_2$, we investigate how much encoded classical information is extractable if, in the first case, $L_1$ and $L_2$ can make global measurements together, and in the second case, $L_1$ and $L_2$ are distant but can perform local operations and classical communication (LOCC). Specifically, in the second case, one-way LOCC, denoted by $\texttt{LOCC}_1$, is being considered.

Our dense coding protocol with non-classical routing can be interpreted as a form of distributed encoding, where the information is coherently delocalized across multiple laboratories. Such a mechanism enlarges the accessible state space, leading to enhanced distinguishability among the encoded states (i.e. a larger set of nearly orthogonal states), and consequently enabling a higher amount of extractable information compared to scenarios where the encoding is localized to a single subsystem.

We use the terminology ``non-classical routing'' to specifically mean that the destination of Alice’s quantum system is placed in quantum superposition and is coherently controlled by an ancillary system. The encoding and the sending cannot be treated as independent steps; rather, they are jointly controlled by the same quantum degree of freedom. This coherent control over the communication path, leading to interference between different routing processes, is the fundamental operational feature of the protocol and the source of the observed advantage.

As we will demonstrate, our protocol provides an advantage over the standard dense coding scheme in certain parameter regimes; however, this advantage is not universal. In particular, for some choices of the shared entangled state, the standard dense coding protocol can perform comparably or even outperform our scheme. This observation motivates a systematic investigation of different scenarios, which we undertake in the following sections. To provide a clear overview, we summarize the regimes in which our protocol exhibits an advantage, as well as those in which it does not, in Table~\ref{tab:summary}.

\subsection{Global Decoding}
Our aim is to obtain the maximum extractable information from the ensemble, $\mathcal{E}$, when Alice non-classically routes her part of the encoded state to the two receivers, $B_1$ and $B_2$. Here, the maximisation is over unitary encodings and the measurements performed by the two receivers.
The Holevo bound maximises the extractable information over all measurements. We then maximise the Holevo bound over all encodings relevant to our protocol. Therefore, the dense coding capacity with non-classical routing is defined as 
\begin{eqnarray}
    \chi_{NCR}^{glo} 
    &=& \max_{|\mathcal{X}|}  \max_{U_2^x,V_2^x,p_X(x)} \chi(\mathcal{E}) \nonumber \\
    &=& \max_{|\mathcal{X}|}  \max_{U_2^x,V_2^x,p_X(x)} \left(S(\rho_{\mathcal{E}}) - \sum_{x \in \mathcal{X}} p_X(x) S(\rho_x) \right) \nonumber \\
     &=& \max_{|\mathcal{X}|} \widetilde{\chi}_{NCR}^{glo} , \;\;\;\;\;\;\;
\end{eqnarray}
where the ensemble is $\mathcal{E} = \left\{ p_X(x), \rho_x \right\}$, the random variable encoded by Alice is $X$, whose realisations $x$ are \textit{letters} in an \textit{alphabet} $\mathcal{X}$, and $S(\rho_x)$ denotes the von Neumann entropy of the state $\rho_x$. Here, $|\mathcal{X}|$ is the size of the alphabet, and $\widetilde{\chi}_{NCR}^{glo}$ is the maximum extractable information for a given alphabet size. The subscript $NCR$ denotes that the scenario considered is in the presence of non-classical routing. We parameterise the ensemble $\mathcal{E}$ by single-qubit unitaries $U_2^x$ and $V_4^x$ (one of which is inclusive of global phase), and the probabilities $p_X(x)$. We choose the initial state $\rho_{B_1AB_2}$ from a few common classes of states. Numerical optimisation is carried out using the \texttt{trust-constr} optimisation algorithm, a method well suited for large-scale optimisations~\cite{conn2000trust}. We analyse and compare our protocol with the standard dense coding protocol with two receivers.

\subsubsection{Pure Entangled States}
\begin{figure}[h!]
\centering
\includegraphics[width=\linewidth]{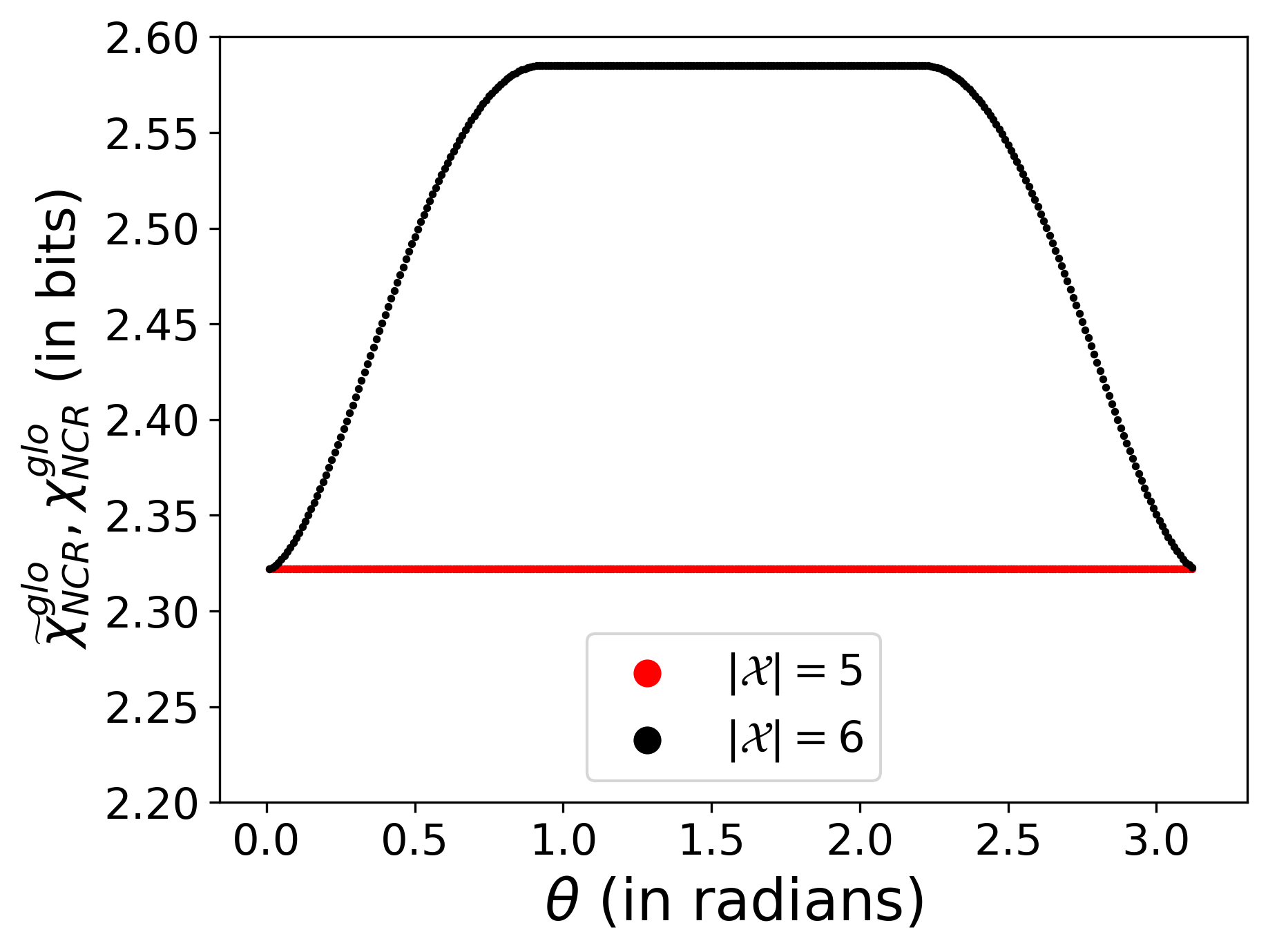}
\caption{The black curve depicts the global dense coding capacity, $\chi_{NCR}^{glo}$, along the vertical axis, as a function of the parameter $\theta$ of the generalised GHZ state, $\ket{gGHZ}$, along the horizontal axis. Intriguingly, we find that other than GHZ states, non-maximally entangled generalised GHZ states also provide a dense coding capacity equal to $\log6$.
The red curve corresponds to the maximum globally extractable information, $\widetilde{\chi}_{NCR}^{glo}$, for an alphabet size $|\mathcal{X}| = 5$. In both cases, the capacities abruptly drop to $\log3$ at values of $\theta$ equal to $0$ and $\pi$, where they become unentangled.} 
\label{fig:GHZ_DC_theta}
\end{figure}

Consider the state shared by Alice, Bob$_1$ and Bob$_2$ to be $\rho_{AB_1B_2} = \outpr{GHZ}{GHZ}_{AB_1B_2}$ where $\ket{GHZ}_{AB_1B_2} = \frac{1}{\sqrt{2}} (\ket{000} + \ket{111})$.  
We find that utilising the three-qubit GHZ state as an input and in the presence of non-classical routing, the maximum extractable information, $\widetilde{\chi}_{NCR}^{glo}$ is equal to $\log{|\mathcal{X}|}$ bits, for $|\mathcal{X}| \le 6$, where $|\mathcal{X}|$ denotes the size of the alphabet. Whereas if $|\mathcal{X}|$ is greater than 6, then the maximum extractable information saturates to $\log{6}$ bits, i.e.
\begin{equation}
    |\mathcal{X}| \le 6 \Rightarrow  \widetilde{\chi}_{NCR}^{glo} = \log{|\mathcal{X}|} ~ \text{bits}, \;\;\; \text{and}
\end{equation}
\begin{equation}
    |\mathcal{X}| \ge 6 \Rightarrow \chi_{NCR}^{glo} = \log{6} \approx 2.5849 ~ \text{bits}.
\end{equation}
Therefore, $\log{6}$ is the global dense coding capacity with non-classical routing while using GHZ state as input. Interestingly, the dense coding capacity of $\log{6}$ is attainable even with a non-maximally entangled state. If the generalised GHZ state of the form $\ket{gGHZ}_{AB_1B_2} = \cos{(\theta/2)}\ket{000} + \sin{(\theta/2)}\ket{111}$, where $0\le \theta \le \pi$, then we find that $\chi_{NCR}^{glo} = \log{6}$ bits is reachable within a wide range of $\theta$, as evident in Fig.~\ref{fig:GHZ_DC_theta}. 

The constancy of the red line with respect to $\theta$ arises from the fact that the corresponding information transmission rate does not depend on the specific value of $\theta$, provided the initial shared generalized GHZ state remains entangled. This result follows from numerical optimization, which shows that for any non-zero amount of entanglement, there exists an optimal encoding along with a corresponding global decoding strategy that enables the reliable transmission of $\log 5$ bits of information. Consequently, the achievable rate remains invariant under variations of $\theta$, which explains the observed constant behavior of the red curve.

In our protocol, the outcome of the measurement performed on the control qubit is to be communicated to the receivers via a noiseless classical channel. Since our protocol makes use of such a classical channel of limited capacity (one classical bit per quantum channel usage), we need to give a boost of an extra bit to the standard dense coding capacity while comparing our protocol with the standard dense coding protocol. Thus, Alice sends three bits of useful information, where two bits  
is due to the pre-shared entanglement in the standard dense coding and one bit due to the access to a noiseless classical channel with limited capacity. Clearly, in scenarios with maximally entangled states, the standard dense coding protocol with three bits is more useful than our dense coding protocol with $\log{6}$ bits. However, two bits of maximal extractable information from the standard dense coding is possible only with a GHZ state, but our protocol has shown maximum ``dense codeability" even with non-maximally entangled generalised GHZ states. Therefore, the trade-off, $\Delta \equiv \chi_{NCR}^{glo} - (\chi_{SDC}+1)$, is the quantity of interest, and $\Delta>0$ implies an advantage in our protocol while $\Delta \le 0$ implies the standard dense coding protocol to be more advantageous. Here $\chi_{SDC}$ refers to the standard dense coding capacity with two receivers. 

\begin{figure}[h!]
\centering
\includegraphics[width=\linewidth]{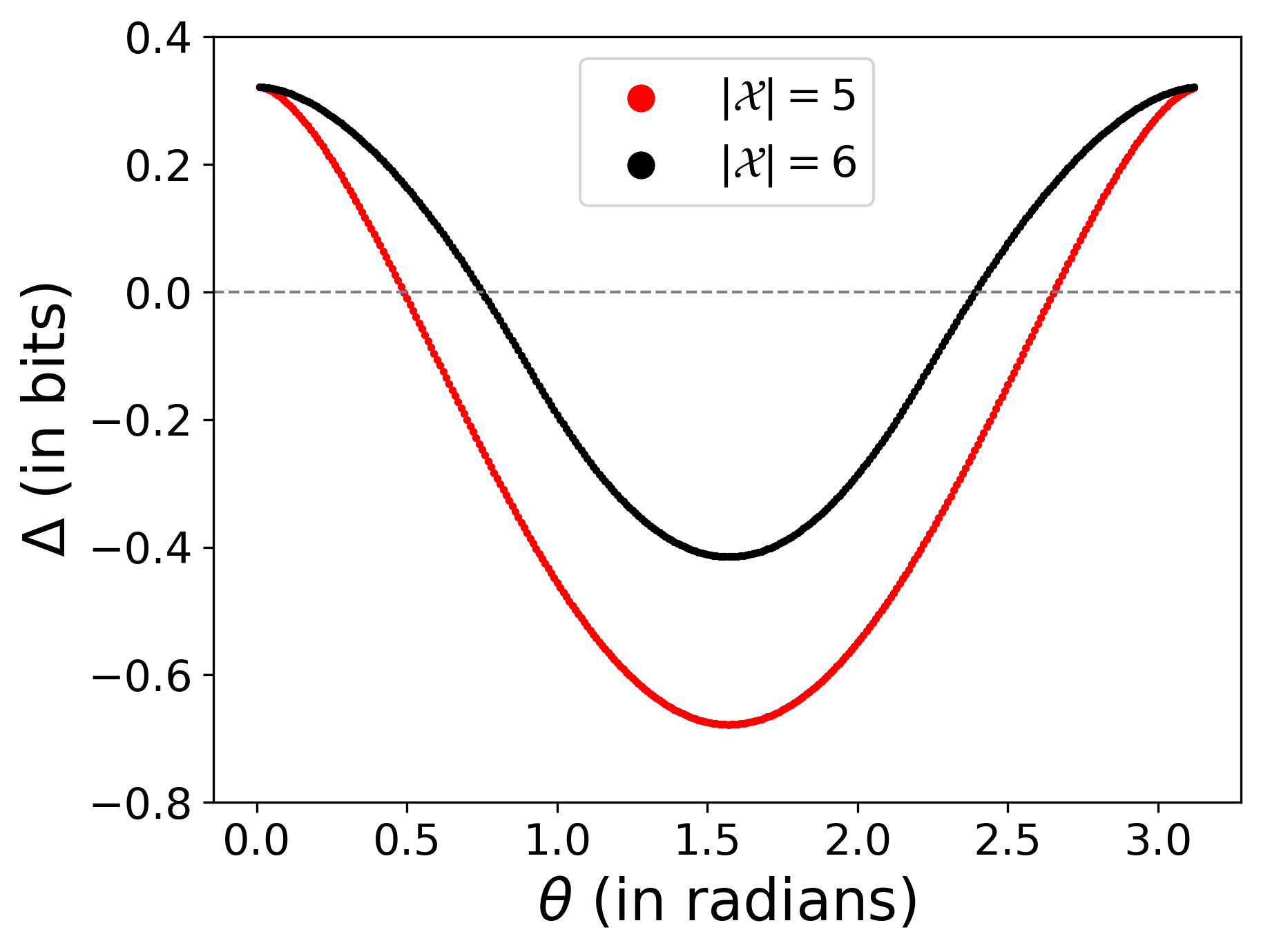}
\caption{The figure of merit chosen for comparing our dense coding protocol with the standard one, given by $\Delta \equiv \chi_{NCR}^{glo} - (\chi_{SDC}+1)$, is plotted along the vertical axis, as a function of $\theta$ of generalised GHZ along the horizontal axis. The black corresponds to our dense coding protocol with  $|\mathcal{X}| = 6$, and the red corresponds to the same with $|\mathcal{X}| = 5$. For low entanglement regions, we find $\Delta>0$, where our protocol shows an advantage over the standard one.}
\label{fig:GHZ_DC_adv}
\end{figure}

In Fig.~\ref{fig:GHZ_DC_adv}, we depict the figure of merit, $\Delta$, as a function of the input state parameter $\theta$. We find that $\Delta>0$ in the range of $\theta \in (0, 0.75] \cup [2.4, \pi)$. Hence, our dense coding protocol performs better than the standard protocol for low-entangled generalised GHZ states. 
This observation is particularly significant because the advantage arises in the regime where the shared entanglement is weak. In practical settings, highly entangled resource states may not always be available or may be difficult to preserve in the presence of noise. Consequently, protocols that retain an advantage in the low-entanglement regime are of considerable interest. Moreover, since the standard dense coding capacity is already optimized within the conventional dense-coding framework, the observed improvement indicates that a quantum superposition of communication routes can provide a communication advantage beyond that achievable through conventional routing strategies.

\subsubsection{Maximally Mixed State}
Let the state shared by $A$, $B_1$ and $B_2$ be $\rho_{AB_1B_2} = \frac{1}{8}(\mathcal{I}_{A} \otimes \mathcal{I}_{B_1} \otimes \mathcal{I}_{B_2})$ where $\mathcal{I}$ is the $2\times2$ Identity operator. 
In an optimal setting, i.e. considering optimal measurements and unitary encodings, there exists an encoding and a decoding scheme where, in the asymptotic limit, we obtain the maximum extractable information and the dense coding capacity, respectively, given by
\begin{equation}
    |\mathcal{X}|\le2 \Rightarrow \widetilde{\chi}_{NCR}^{glo}  = \log{|\mathcal{X}|} ~ \text{bits}, \;\;\; \text{and}
\end{equation}
\begin{equation}
    |\mathcal{X}| \ge 3 \Rightarrow  \chi_{NCR}^{glo} \approx 1.2539 ~\text{bits}
\end{equation}
respectively, where $|\mathcal{X}|$ is the size of the alphabet corresponding to the random variable that Alice encodes. For a maximally mixed state, the standard dense coding capacity is zero, and therefore $\Delta \approx 0.2539$. This suggests that non-classical routing is able to utilise a maximally mixed state for dense coding. 

\subsubsection{Completely Separable Mixed State}
Let the initial state shared among the three parties be a completely separable mixed one, given by
\begin{equation}
\label{mixed}
    \rho_{AB_1B_2} = \sum_{i=1}^{k} p_i \rho_A^i \otimes \rho_{B_1}^i \otimes \rho_{B_2}^i,
\end{equation}
where, for our purposes, each of the three subsystems is a qubit. We consider $\rho^i_{A(B_1,B_2)} \equiv \frac{1}{2}(\mathcal{I} + \hat{n}_{A(B_1,B_2)}\cdot\vec{\sigma})$, where $\hat{n}_{A(B_1,B_2)}$ is a unit vector, $\vec{\sigma}$ is the Pauli vector, and $p_i\ge 0$, $\forall i$, with $\sum_i p_i = 1$. 

We find that, for $k=2$,  and in an optimal setting, there exists an encoding and a decoding scheme for which the maximum extractable information for $|\mathcal{X}|\le 4 $, and $|\mathcal{X}| = 5$ are given by
\begin{equation}
    |\mathcal{X}|\le 4 \Rightarrow \widetilde{\chi}_{NCR}^{glo}  = \log{|\mathcal{X}|} ~ \text{bits}, \;\;\; \text{and}
\end{equation}
\begin{equation}
    |\mathcal{X}| = 5 \Rightarrow \widetilde{\chi}_{NCR}^{glo} \approx \log{5} ~\text{bits}
\end{equation}
respectively. We also found a completely separable mixed state of purity of $0.52$ with $\widetilde{\chi}_{NCR}^{glo} \approx \log{5}$. It is numerically calculated that for this state, $\chi_{SDC} = 1$, therefore giving us $\Delta \approx \log{5/4}>0$ bits. Therefore, utilising non-classical routing demonstrates that classical correlation helps in a dense coding protocol.

Pure states, which are unentangled in the bipartition of sender to receivers, prove to be inefficient for the dense coding protocol using non-classical routing. If the input state is $\rho_{AB_1B_2} = \outpr{\psi}{\psi}$ such that $\ket{\psi}_{AB_1B_2} \equiv \ket{\phi}_A \otimes \ket{\Phi}_{B_1B_2}$, then, an observation is made that in an optimal setting, there exists an encoding and a decoding scheme where, in the asymptotic limit, we find that
\begin{equation}
    |\mathcal{X}|\le 3 \Rightarrow \widetilde{\chi}_{NCR}^{glo}  = \log{|\mathcal{X}|} ~ \text{bits}, \;\;\; \text{and}
\end{equation}
\begin{equation}
    |\mathcal{X}| \ge 3 \Rightarrow  \chi_{NCR}^{glo} = \log{3} \approx 1.5849 ~\text{bits}.
\end{equation}
In this scenario, $\chi_{SDC} = 1$ as $\rho_{AB_1B_2}$ is a pure product state, but our protocol gives $\chi_{NCR}^{glo} = \log{3}$, therefore $\Delta = \log{\frac{3}{4}} < 0$. Thus our protocol does not perform better than the standard protocol using pure states which are unentangled in the bipartition of sender to receivers.

In the above analyses, since we perform non-linear numerical optimisation, the values above are, in fact, lower bounds to $\chi_{NCR}^{glo}$.

\subsection{One-way LOCC Decoding}
In the global decoding scenario, the receiver labs, $L_1$ and $L_2$, can be assumed to be a single lab $L$ as they are close enough to implement global measurements. However, there might arise situations where the two labs are distant from each other, do not have access to global measurements, and can only perform LOCC-type measurements to extract information. 

The maximal information that is extractable from a bipartite ensemble restricted to local operations and classical communication is referred to as the locally accessible information, $I_{acc}^{LOCC}$, of that ensemble~\cite{xn--Badziag-npe2003Sep}. Like globally accessible information, there are no methods to calculate locally accessible information analytically, and it is even more complex because the set of LOCC measurements has no tractable characterisation~\cite{Chitambar2014May}. Nevertheless, an upper bound to locally accessible information was provided in~\cite{xn--Badziag-npe2003Sep}. This bound is given by
\begin{equation}
    I_{acc}^{LOCC} \le S(\rho_A) + S(\rho_B) - \max_{Z=A,B} \sum_{x \in \mathcal{X}} p_X(x) S(\rho_Z^x)
\end{equation}
where $\mathcal{E} = \{ p_X(x), \rho_{AB}^x \}$ is the bipartite ensemble, $\rho_A$ and $\rho_B$ are reductions of $\rho_{\mathcal{E}} = \sum_x p_X(x) \rho_{AB}^x$, and $\rho_Z^x$ is a reduction of $\rho_{AB}^x$, and $S(\rho_Z^x)$ denotes the von Neumann entropy of $\rho_Z^x$. A slightly stronger bound and a lower bound have also been looked into~\cite{Horodecki2004Oct, Sen(De)2006Nov}. Unfortunately, unlike the Holevo bound, this Holevo-like bound cannot be universally saturated for all ensembles, even in the asymptotic limit. However, some examples of ensembles do saturate it in $\mathbb{C}^n \otimes \mathbb{C}^n$ systems~\cite{xn--Badziag-npe2003Sep}. 

We approach this problem by relaxing the measurements to the class of one-way LOCC ones. We denote such one-way LOCC measurement protocols by $\texttt{LOCC}_1$. The ensemble $\mathcal{E} = \{p_X(x), \rho_x^{L_1L_2}\}$ is with labs $L_1$ and $L_2$. Without loss of generality, Bob$_1$ starts the $\texttt{LOCC}_1$ protocol by performing a positive operator valued measurement (POVM) $\mathcal{M} = \{\Lambda_{y_1}^{L_1}\}_{y_1 \in \mathcal{Y}_1}$ on his subsystem (lab $L_1$) of the ensemble. After measurement, he obtains an outcome $y_1^* \in \mathcal{Y}_1$ with probability $p_{Y_1}(y_1^*)$. Here, $Y_1$ is the information extracted by Bob$_1$ in the form of a random variable whose realisations $y_1$ are letters in the alphabet $\mathcal{Y}_1$. The outcome $y_1^*$ is then classically communicated to Bob$_2$ by Bob$_1$. Depending on this information received from Bob$_1$, Bob$_2$ performs the measurement $\mathcal{M}_2^{y_1^*} \equiv \{ \Lambda_{y_2|y_1^*}^{L_2}\}_{y_2 \in \mathcal{Y}_2}$ on the post-measurement ensemble and obtains outcome $y_2^* \in \mathcal{Y}_2$. The $\texttt{LOCC}_1$ protocol ends here, and the total information extracted by both the receivers is given by $I(X;Y_1Y_2)$. Using the chain rule~\cite{wilde2013quantum}, the quantity, $I(X;Y_1Y_2)$, simplifies to
\begin{multline}
    I(X;Y_1Y_2) = I(X; Y_1) + I(X;Y_2|Y_1) \\ = I(X; Y_1) + \sum_{y_1} p_{Y_1}(y_1) I(X;Y_2|Y_1=y_1),
\end{multline}
where $\Lambda^{L_1} \equiv \Lambda^{L_1} \otimes \mathcal{I}^{L_2}$, $\Lambda^{L_2} \equiv \mathcal{I}^{L_1} \otimes \Lambda^{L_2}$, $p_{Y_1}(y_1) = \tr\left\{\Lambda_{y_1}^{L_1} \rho_{\mathcal{E}}\right\}$ and $\rho_{\mathcal{E}}$ is the density matrix associated with ensemble $\mathcal{E}$.

After Bob$_1$ performs a measurement on the state, $\rho_x^{L_1L_2}$, the encoded ensemble, $\mathcal{E} = \{p_X(x), \rho_x^{L_1L_2}\}$ collapses to the post-measurement ensemble $\mathcal{E}^{y_1^*} = \{p_{X|Y}(x|y_1^*), \rho_{x|y_1^*}^{L_1L_2}\}$, yielding an outcome $y_1^*$,  where the post-measurement ensemble  takes the explicit form, given by
\begin{equation}
    \Bigg\{ \frac{p_X(x) \tr\{\Lambda_{y_1^*}^{L_1} \rho_x^{L_1L_2}\}}{ \sum_x p_X(x) \tr\{\Lambda_{y_1^*}^{L_1} \rho_x^{L_1L_2}\}}, \frac{\sqrt{\Lambda_{y_1^*}^{L_1}} \rho_x^{L_1L_2} \sqrt{\Lambda_{y_1^*}^{L_1}}}{\tr \{\Lambda_{y_1^*}^{L_1} \rho_x^{L_1L_2} \} } \Bigg\}.
\end{equation}
The encoded ensemble $\mathcal{E}$ is referred to calculate $I(X;Y_1)$, and the post-measurement ensemble $\mathcal{E}^{y_1^*}$ is referred to calculate $I(X;Y_2|Y_1 = y_1^*)$. For a convenient numerical analysis, we restrict the set of POVMs to rank-one projective measurements. The four-dimensional rank-one projective measurement set is given by $\{\Pi_i^{L_1} \equiv U\outpr{i}{i}U^{\dagger}\}_{i=1}^4$ where $\ket{i}$ are two-qubit basis states and $U$ is an arbitrary two-qubit unitary operator of the form
\begin{equation}
    U = (A \otimes B)U_d(C \otimes D),
\end{equation}
where $A, B, C, D \in U(2)$. $U_d$ is a ``non-local" unitary operator belonging to $\mathbb{C}^2 \otimes \mathbb{C}^2$, given by
\begin{equation}
    U_d = \exp(-i[\alpha_x \sigma_x \otimes \sigma_x + \alpha_y \sigma_y \otimes \sigma_y + \alpha_z \sigma_z \otimes \sigma_z])
\end{equation}
where $\alpha_x, \alpha_y, \alpha_z \in \mathbb{R}$~\cite{uni1,uni2,uni3}. We optimise the extractable information, $I(X;Y_1Y_2)$, over rank-one projective measurements, $\Lambda_{y_1^*}^{L_1}=\Pi_{y_1^*}^{L_1}$, probabilities $p_X(x)$ and all encoding unitaries, using \texttt{trust-constr} optimisation method. 

Considering a generalised GHZ state as $\rho_{AB_1B_2}$, the maximum extractable information by the two receivers, if they have access to only one-way LOCC protocols, is given by
\begin{eqnarray}
   I(X;Y_1Y_2) &=& 2.3215~ \text{bits}, 
\end{eqnarray}
which suggests $I_{acc}^{LOCC_1} \ge 2.3215 ~ \text{bits}$, corresponding to alphabet size $ |\mathcal{X}| = 5 $. This occurs when the generalised GHZ state has negligible but non-zero entanglement as $\theta \approx 0.01470$ and $\theta \approx 6.26848$, both around $0.01470$ radians away from $\theta=0$ and $\theta=2\pi$, respectively. Interestingly, we also find that there exist one-shot $\texttt{LOCC}_1$ decoding protocols that saturate $\widetilde{\chi}_{NCR}^{glo}$ for $|\mathcal{X}| \le 4$, i.e, 
\begin{equation}
    |\mathcal{X}| \le 4 \Rightarrow I(X;Y_1Y_2) = \log{|\mathcal{X}|} = \widetilde{\chi}_{NCR}^{glo}.
\end{equation}
We compare our one-way LOCC decoding protocol with the standard LOCC decoding protocol. However, to keep the comparison on equal footing, the equivalent setting in the standard case to be considered is that Alice sends her encoded quantum system to Lab$_1$ with probability $p$ and Lab$_2$ with probability $1-p$. The optimal $\rho_{AB_1B_2}$ obtained for the one-shot $\texttt{LOCC}_1$ decoding protocol is a generalised GHZ state with $\theta \approx 0.01470$. If one uses such a generalised GHZ state, then the global dense coding capacity for the standard protocol is $\chi_{SDC} \approx 1.00084$. Therefore, the figure of merit, $\widetilde{\Delta} = I(X;Y_1Y_2) - (\chi_{SDC}+1) \ge  2.3215 - (1.0008+1) = 0.3206$, implying advantage even in our $\texttt{LOCC}_1$-based decoding protocol over the standard dense coding scenario with global decoding.

\section{Noisy Dense Coding with Non-classical Routing} ~\label{sec:noise}

In Section \ref{sec:ourprotocol}, we restricted our analysis to ideal resources, assuming a noiseless classical channel for communicating measurement outcomes and coherent control over noiseless channels. In this section, we examine how the presence of noise in these resources affects the protocol.

\subsection{Quantum Noise in the Protocol}
In the noiseless setting, each process corresponds to Alice applying a unitary encoding followed by transmission of her system through a noiseless quantum channel to the appropriate laboratory. Consequently, the state undergoes a unitary transformation before reaching the corresponding lab $L_i$. 
When Processes 1 and 2 are coherently controlled, the control system determines which of the two processes is applied. Prior to any measurement on the control, the joint control-target state evolves under the corresponding multiplexed unitary transformation. For an initial control state $\ket{+}$, the overall transformation takes the form
\begin{multline}
    \ket{\psi}_{L_1L_2} \otimes \ket{+} 
    \longrightarrow 
    \frac{1}{\sqrt{2}}\Big(
    \ket{0}_c \otimes U^x_2 \ket{\psi}_{L_1L_2} \nonumber \\ 
    + 
    \ket{1}_c \otimes V^x_4 \cdot \texttt{SWAP}_{2,4} \ket{\psi}_{L_1L_2}
    \Big),
\end{multline}
where the laboratories are treated as \textit{ordered} quantum systems $L_1 \equiv (B_1,A)$ and $L_2 \equiv (B_2,C_{B_2})$.

We can define a noisy scenario to be such that after the encoding operation, the transmitted system is sent through a noisy quantum channel. More precisely, in Process $i$ the system is acted upon by a completely positive trace-preserving (CPTP) map $\mathcal{E}_i$ before reaching the corresponding laboratory $L_i$, instead of being transmitted through the identity channel as in the noiseless case.
While in the noiseless scenario coherent control was required only over unitary operations, extending this notion to general CPTP maps is not straightforward. To address this, we follow the approach of~\cite{Abbott2020Sep} and implement coherent control over the channels by coherently controlling their Stinespring dilation unitaries.

To incorporate noise into our protocol, it is useful to first understand how two CPTP maps can be coherently superposed. Let $\mathcal{E}_1$ and $\mathcal{E}_2$ be two CPTP maps characterized by the Kraus operators $\{K_i\}_i$ and $\{L_j\}_j$, respectively. Their Stinespring dilations can be expressed as
\begin{equation}
    \mathcal{E}_1:~\ket{\psi}_t \otimes \ket{\epsilon_1}_{E_1} \longrightarrow \sum_i K_i \ket{\psi}_t \otimes \ket{i}_{E_1},
\end{equation}
\begin{equation}
    \mathcal{E}_2:~\ket{\psi}_t \otimes \ket{\epsilon_2}_{E_2} \longrightarrow \sum_j L_j \ket{\psi}_t \otimes \ket{j}_{E_2},
\end{equation}
where $E_1$ and $E_2$ denote two initially uncorrelated environments prepared in states $\ket{\epsilon_1}_{E_1}$ and $\ket{\epsilon_2}_{E_2}$, respectively.

Introducing a control qubit initialized in the state $\ket{+}_c$, the joint control-target-environment system evolves unitarily according to
\begin{multline}
\ket{+}_c\otimes\ket{\psi}_t\otimes\ket{\epsilon_1}_{E_1}\otimes\ket{\epsilon_2}_{E_2}
\longrightarrow \\
\frac{1}{\sqrt{2}}\bigg(
\ket{0}_c\otimes 
\left[\sum_i K_i\ket{\psi}_t\otimes\ket{i}_{E_1}\right]\otimes\ket{\epsilon_2}_{E_2} \\
+ 
\ket{1}_c\otimes 
\left[\sum_j L_j\ket{\psi}_t\otimes\ket{j}_{E_2}\right]\otimes\ket{\epsilon_1}_{E_1}
\bigg).
\end{multline}
Thus, the control qubit coherently selects between the Stinespring unitaries implementing the two CPTP maps, acting jointly on the target system and the corresponding environments $E_1$ and $E_2$. A superposition of the two channels is obtained by subsequently measuring the control qubit in the Hadamard basis.

The noiseless coherent routing protocol can be extended to the noisy setting using the approach described above. Let $\mathcal{E}_1$ and $\mathcal{E}_2$ be two CPTP maps with Kraus operators 
$\{\mathcal{I}_{B_1}\otimes K^i \otimes \mathcal{I}_{B_2}\otimes \mathcal{I}_{C_{B_2}}\}_i \equiv \{K^i_2\}_i$ 
and 
$\{\mathcal{I}_{B_1}\otimes \mathcal{I}_{C_{B_1}}\otimes \mathcal{I}_{B_2}\otimes L^j\}_j \equiv \{L^j_4\}_j$, respectively.
Following the Stinespring dilation picture, the corresponding processes can be coherently controlled by introducing environments $E_1$ and $E_2$ and allowing the control qubit to coherently select between the two dilation unitaries. In our protocol, the joint control–target–environment system then evolves as
\begin{widetext}
\begin{multline}
\ket{+}_c \ket{\psi}_{L_1L_2} \ket{\epsilon_1}_{E_1} \ket{\epsilon_2}_{E_2}
\longrightarrow 
\frac{1}{\sqrt{2}} \bigg[
\ket{0}_c \otimes \ket{\epsilon_2}_{E_2} \otimes 
\bigg( \sum_i K^i_2 U^x_2 \ket{\psi}_{L_1L_2} \otimes \ket{i}_{E_1} \bigg)
\\
+
\ket{1}_c \otimes \ket{\epsilon_1}_{E_1} \otimes 
\bigg( \sum_j L^j_4 V^x_4 \cdot \texttt{SWAP}_{2,4}\ket{\psi}_{L_1L_2} \otimes \ket{j}_{E_2} \bigg)
\bigg].
\end{multline}
\end{widetext}
Here $\{U^x, V^x\}$ denote the pair of encoding unitaries applied by Alice for the classical letter $x$. The subscripts $2$ and $4$ indicate that the corresponding operators act on the second and fourth subsystems of the ordered state $\ket{\psi}_{L_1L_2}$.

The superposition is realized by measuring the control qubit in the Hadamard basis. Conditioned on obtaining the outcome corresponding to $\ket{+}$, the (unnormalized) post-measurement state is given by
\begin{multline}
\ket{\phi} =
\ket{\epsilon_2}_{E_2} \otimes \sum_i M_{ix}\ket{\psi}_{L_1L_2}\ket{i}_{E_1}
\\
+
\ket{\epsilon_1}_{E_1} \otimes \sum_j N_{jx}\ket{\psi}_{L_1L_2}\ket{j}_{E_2},
\end{multline}
where $M_{ix} = K^i_2 U^x_2$ and $N_{jx} = L^j_4 V^x_4 \cdot \texttt{SWAP}_{2,4}$. Tracing out the environments yields the final state shared by the receivers,
\begin{multline}
\rho_x = \frac{1}{\tr\{\rho_x\}} \bigg(
\mathcal{E}_1(\mathcal{U}_x(\rho_t))
+
\mathcal{E}_2(\mathcal{V}_x(\mathcal{S}(\rho_t)))
\\
+
T_2 \rho_t T_1^\dagger
+
T_1 \rho_t T_2^\dagger
\bigg),
\end{multline}
where $\rho_t = \rho_{B_1AB_2} \otimes \outpr{0}{0}$, and $\mathcal{U}_x$, $\mathcal{V}_x$, and $\mathcal{S}$ denote unitary channels generated by $\{U^x_2\}$, $\{V^x_4\}$, and $\{\texttt{SWAP}_{2,4}\}$, respectively. The operators $T_1 = \sum_i \langle \epsilon_1 | i \rangle M_{ix}$ and $T_2 = \sum_j \langle \epsilon_2 | j \rangle N_{jx}$ are known as transformation matrices~\cite{Abbott2020Sep}.

This procedure generates an encoded ensemble $\mathcal{E} \equiv \{p_X(x), \rho_x\}$. Equivalently, the resulting state can be expressed in terms of a noisy multiplexed unitary acting on the extended system. Let $\rho_{tE_1E_2c} = \rho_t \otimes \rho_{E_1} \otimes \rho_{E_2} \otimes \rho_c$, where $\rho_{E_1} = \outpr{\epsilon_1}{\epsilon_1}$, $\rho_{E_2} = \outpr{\epsilon_2}{\epsilon_2}$, and $\rho_c = \outpr{+}{+}$. Then,
\begin{equation}
\tr_{cE_1E_2}\left[
\frac{
(\mathcal{I}_{tE_1E_2} \otimes \Pi_{+})
\, W \rho_{tE_1E_2c} W^\dagger \,
(\mathcal{I}_{tE_1E_2} \otimes \Pi_{+})
}{
\tr\left[
(\mathcal{I}_{tE_1E_2} \otimes \Pi_{+})
\, W \rho_{tE_1E_2c} W^\dagger
\right]
}
\right],
\end{equation}
where the effective noisy multiplexing unitary $W$ is given by
\begin{multline}
W =
\outpr{0}{0}_c \otimes \mathcal{I}_{E_2} \otimes
\big(\mathcal{I}_{B_1} \otimes [U_{\mathcal{E}_1} \circ (U^x \otimes \mathcal{I}_{E_1})] \otimes \mathcal{I}_{B_2} \otimes \mathcal{I}_{C_{B_2}} \big)
\\
+
\outpr{1}{1}_c \otimes \mathcal{I}_{E_1} \otimes
\big(\mathcal{I}_{B_1} \otimes \mathcal{I}_{C_{B_1}} \otimes \mathcal{I}_{B_2} \otimes [U_{\mathcal{E}_2} \circ (V^x \otimes \mathcal{I}_{E_2})] \big)
\cdot \texttt{SWAP}_{2,4},
\end{multline}
with $U_{\mathcal{E}_1}$ and $U_{\mathcal{E}_2}$ denoting Stinespring dilation unitaries of the channels $\mathcal{E}_1$ and $\mathcal{E}_2$, respectively.

\begin{figure}[h!]
\centering
\includegraphics[width=\linewidth]{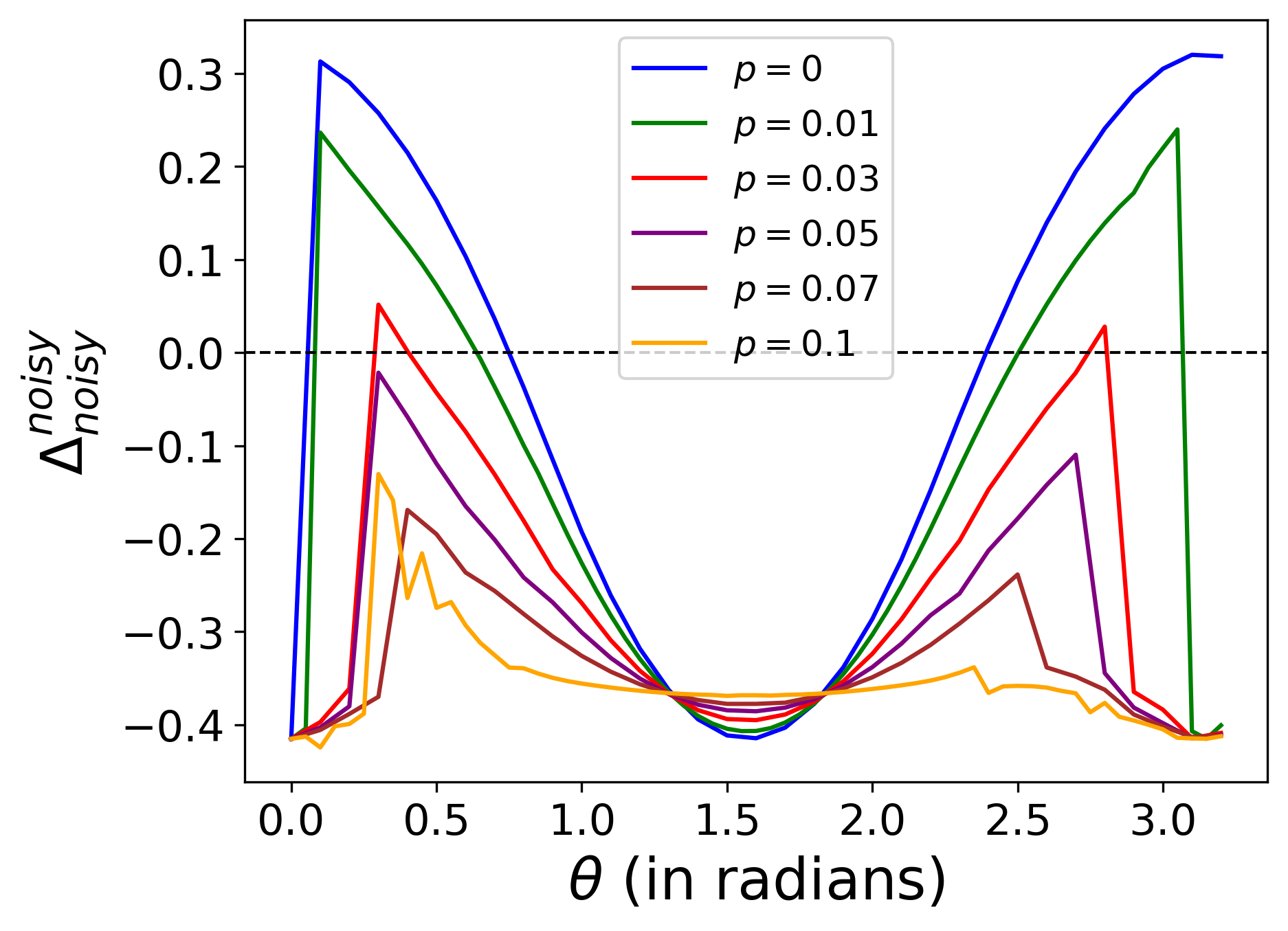}
\caption{The figure of merit chosen for our dense coding protocol with the standard one, both under the same quantum noise, given by $\Delta_{noisy}^{noisy} = \chi_{NCR, N}^{glo}-(\chi_{SDC, N}^{glo}+1)$, is plotted along the vertical axis, as a function of $\theta$ of generalised GHZ along the horizontal axis. 
The reported values correspond to numerically obtained lower bounds on $\Delta_{\mathrm{noisy}}^{\mathrm{noisy}}$. For low entanglement regions and dephasing probability less than $0.03$, we find $\Delta_{noisy}^{noisy}>0$, where our protocol shows an advantage over the standard one, both under the same quantum noise.}
\label{fig:noisy_noisy}
\end{figure}

\begin{figure}[h!]
\centering
\includegraphics[width=\linewidth]{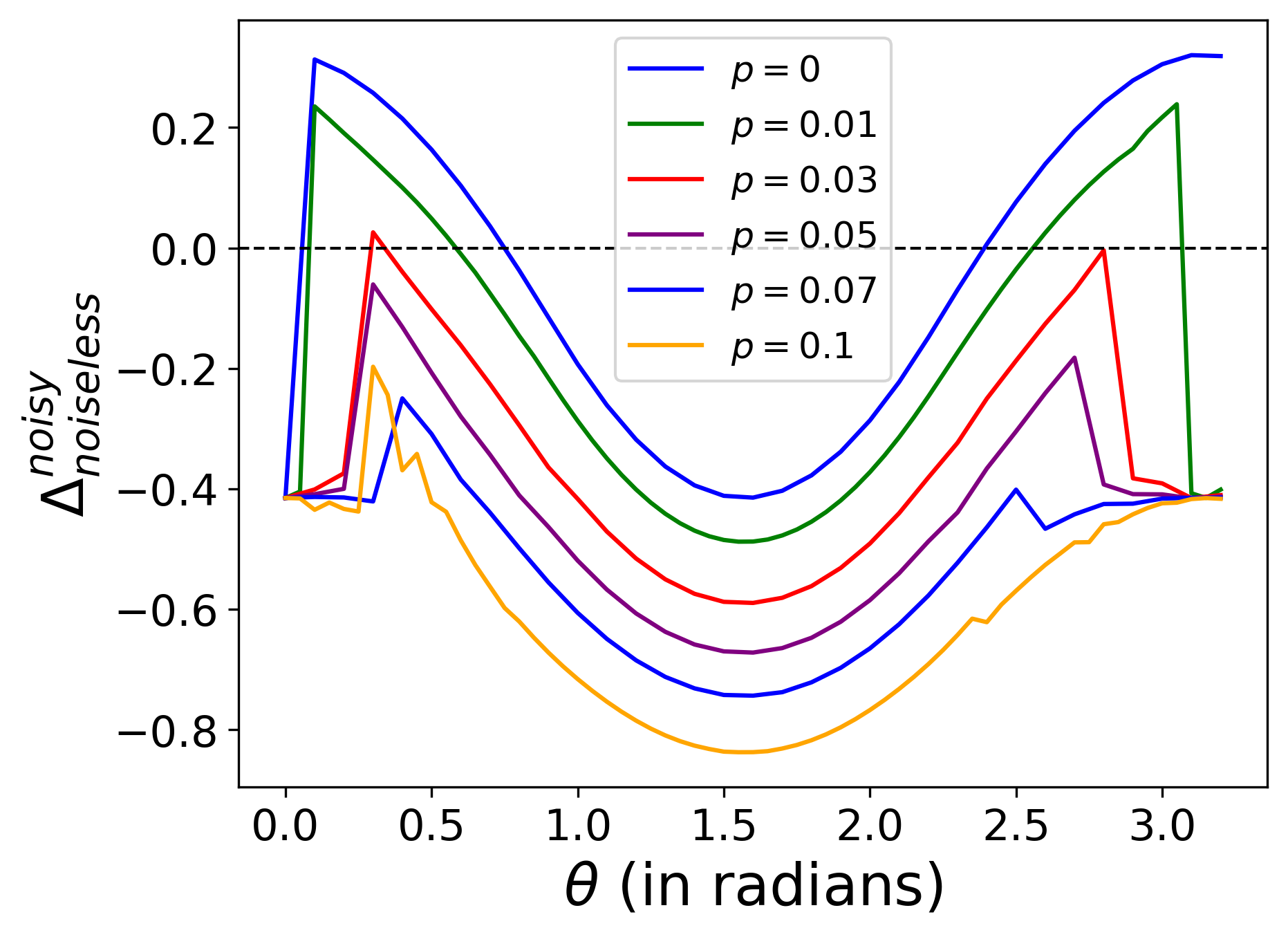}
\caption{The figure of merit chosen for our dense coding protocol with the standard one, both under the same quantum noise, given by $\Delta_{noiseless}^{noisy}>0 = \chi_{NCR, N}^{glo}-(\chi_{SDC}^{glo}+1)$, is plotted along the vertical axis, as a function of $\theta$ of generalised GHZ along the horizontal axis. 
The reported values correspond to numerically obtained lower bounds on $\Delta_{noiseless}^{noisy}>0$. For low entanglement regions and dephasing probability less than $0.03$, we find $\Delta_{noiseless}^{noisy}>0$, where our noisy protocol shows an advantage over even the noiseless standard one.} 
\label{fig:noisy_ideal}
\end{figure}

To quantify the advantage of our protocol, we consider the following figures of merit:
\begin{equation}
\Delta_{\mathrm{noisy}}^{\mathrm{noisy}} 
= \chi_{\mathrm{NCR},N}^{\mathrm{glo}} 
- \big(\chi_{\mathrm{SDC},N}^{\mathrm{glo}} + 1\big),
\end{equation}
\begin{equation}
\Delta_{\mathrm{noiseless}}^{\mathrm{noisy}} 
= \chi_{\mathrm{NCR},N}^{\mathrm{glo}} 
- \big(\chi_{\mathrm{SDC}}^{\mathrm{glo}} + 1\big).
\end{equation}
Here, $\chi_{\mathrm{NCR},N}^{\mathrm{glo}}$ denotes the dense coding capacity of the non-classical routing (NCR) protocol in the presence of quantum noise, while $\chi_{\mathrm{SDC},N}^{\mathrm{glo}}$ represents the capacity of the standard dense coding (SDC) protocol under the same noisy conditions. The quantity $\chi_{\mathrm{SDC}}^{\mathrm{glo}}$ corresponds to the noiseless SDC capacity. The additive constant $1$ accounts for the classical communication cost required in the protocol.
The quantity $\Delta_{\mathrm{noisy}}^{\mathrm{noisy}}$ therefore measures the advantage of the NCR protocol over the standard DC protocol when both are subject to identical quantum noise. In contrast, $\Delta_{\mathrm{noiseless}}^{\mathrm{noisy}}$ compares the performance of the noisy NCR protocol against the ideal (noiseless) SDC protocol, thereby capturing the robustness of our scheme against noise. In our numerical simulations, the optimization underlying $\chi_{\mathrm{NCR},N}^{\mathrm{glo}}$ and $\chi_{\mathrm{SDC},N}^{\mathrm{glo}}$ is performed not only over the encoding operations but also over the choice of environmental states, which are taken to be qubits.

In our analysis, we consider quantum dephasing noise, implemented by the Kraus operators $\{\sqrt{1-p}~\mathcal{I}, \sqrt{p}~\sigma_{z}\}$, where $p$ denotes the dephasing probability. From  Fig.~\ref{fig:noisy_noisy}, we observe that there exist values of the parameter $\theta$ for which $\Delta_{\mathrm{noisy}}^{\mathrm{noisy}} > 0$, indicating a clear advantage of the NCR protocol over its standard counterpart in the noisy regime. Remarkably, we also find from Fig.~\ref{fig:noisy_ideal} that $\Delta_{\mathrm{noiseless}}^{\mathrm{noisy}} > 0$ for dephasing probabilities $p \leq 0.03$, demonstrating that the noisy NCR protocol can outperform even the noiseless standard DC protocol in this parameter regime. 
 \\

\subsection{Classical Communication Noise}
In the preceding subsection, we analyzed the effect of quantum noise acting during the encoding and transmission stages of the dense coding protocol with nonclassical routing. We now turn to the impact of noise in the classical communication channel used to transmit the measurement outcomes.

In the presence of classical noise, reliable communication requires encoding the measurement outcomes using redundancy, so that classical error-correcting techniques can be employed. As a consequence, the effective classical communication cost increases from one classical bit to $1+\epsilon$ bits per use of the protocol, where $\epsilon$ depends on the noise characteristics of the channel and the specific error-correcting scheme employed.

Taking this additional cost into account, the figure of merit is modified to
\begin{equation}
\Delta' = \chi_{\mathrm{NCR}}^{\mathrm{glo}} - \big(\chi_{\mathrm{SDC}}^{\mathrm{glo}} + 1 + \epsilon\big).
\end{equation}
It therefore follows that for sufficiently large $\epsilon$, the advantage offered by the nonclassical routing protocol may diminish or even vanish. \\

\section{Multiple Receivers} \label{sec:distributed}
Consider a single sender, $A$, and $M$ receivers, $B_1, \cdots, B_M$,  where every $B_i$ resides in laboratory $L_i$. The sender and receivers share a $(M+1)$-qubit generalised GHZ state
\begin{equation}
    \ket{gGHZ}_{AB_1\cdots B_M} = \cos{\frac{\theta}{2}}\ket{0}^{\otimes (M+1)} + e^{i\phi}\sin{\frac{\theta}{2}}\ket{1}^{\otimes (M+1)}.
\end{equation}
Each laboratory $L_i$ has access to an auxiliary state $\ket{0}_{C_{B_i}}$.
We can generalise our protocol involving two receivers to $M$ receivers by creating a superposition of $M$ such processes. The control quantum system in such a scenario is an $M$-dimensional quantum state. For a given $x \in \mathcal{X}$, let $\{U^x_i\}_{i=1}^{M}$ be the set of unitaries to be multiplexed. The multiplexed unitary is then given by
\begin{equation}
    W_x = \sum_{i=1}^M (U_i^x)_{2i} \cdot \texttt{SWAP}_{2, 2i} \otimes \outpr{i-1}{i-1}_c,
\end{equation}
where $(U_j^x)_{2j}$ refers to single-qubit unitary $U_j^x \in \{U^x_i\}_{i=1}^{M}$ where all indices other than $2j^{th}$ index are identity operators, and $\texttt{SWAP}_{2, 2i}$ is a $\texttt{SWAP}$ unitary acting on indices 2 and $2i$ ($\texttt{SWAP}_{2, 2}$ is the Identity operator). 

The unitary operator, $W_x$, acts on the joint state of the target and control, viz. $W_x(\rho \otimes \rho_c)W_x^{\dagger}$, where $\rho_c = \outpr{+_d}{+_d}$ with $\ket{+_d} = \frac{1}{\sqrt{d}} \sum_{j=0}^{d-1} \ket{j}$, and  $\rho = \outpr{\psi}{\psi}$ with $\ket{\psi} = \ket{\psi}_{B_1A:B_2C_{B_2}:\cdots:B_MC_{B_M}} \equiv \ket{\psi}_{L_1L_2\cdots L_M}$. After the unitary evolution, a measurement is performed on the control state in the generalised Hadamard basis (for qudits), and the outcome corresponding to $\outpr{+_d}{+_d}$ basis is post-selected. Then, the state now possessed at the receivers' end is given by
\begin{equation}
    \ket{\Psi_x} = \frac{1}{\sqrt{\inpr{\Psi_x}{\Psi_x}}} \left(   \sum_{i=1}^M (U_i^x)_{2i} \cdot \texttt{SWAP}_{2, 2i}   \right) \ket{\psi}_{L_1L_2\cdots L_M},
\end{equation} 
with probability $p_X(x)$. 

Using a $M$-dimensional quantum state as a control system, Alice coherently encodes and routes her quantum system to all the laboratories from $L_1$ to $L_M$, thereby creating the encoded ensemble $\mathcal{E} = \{p_X(x), \outpr{\Psi_x}{\Psi_x}\}$ possessed by all labs together. 

\subsection{Global Decoding} 
We consider two cases, first when the shared state is entangled in the bipartition $A:B_1...B_M$, and second when the state is unentangled in such a bipartition.

\begin{figure}[h!]
\centering
\includegraphics[width=\linewidth]{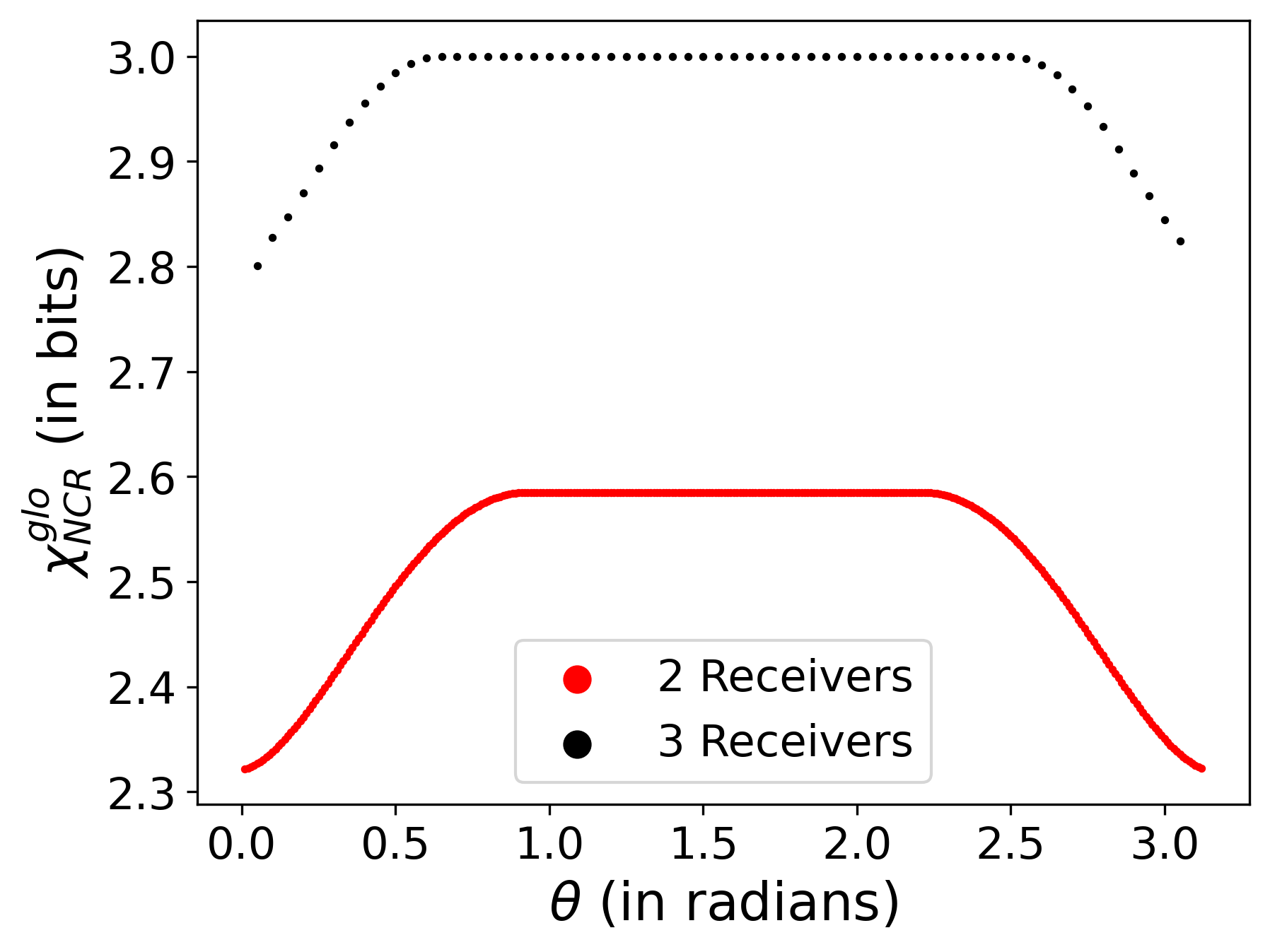}
\caption{The black and red curves depict the global dense coding capacity with non-classical routing, given by $\chi_{DC}^{NCR}$, along the vertical axis, as a function of the parameter $\theta$ of the generalised GHZ state. The red solid curve corresponds to two receivers, while the black dotted curve is for the three-receivers case. 
The dense coding capacity reaches a maximum for a greater range of $\theta$ in $M=3$ case than $M=2$ case.} 
\label{fig:GHZ_DC_theta_3receivers}
\end{figure}

\subsubsection{Entangled Pure States}
\begin{figure}[h!]
\centering
\includegraphics[width=\linewidth]{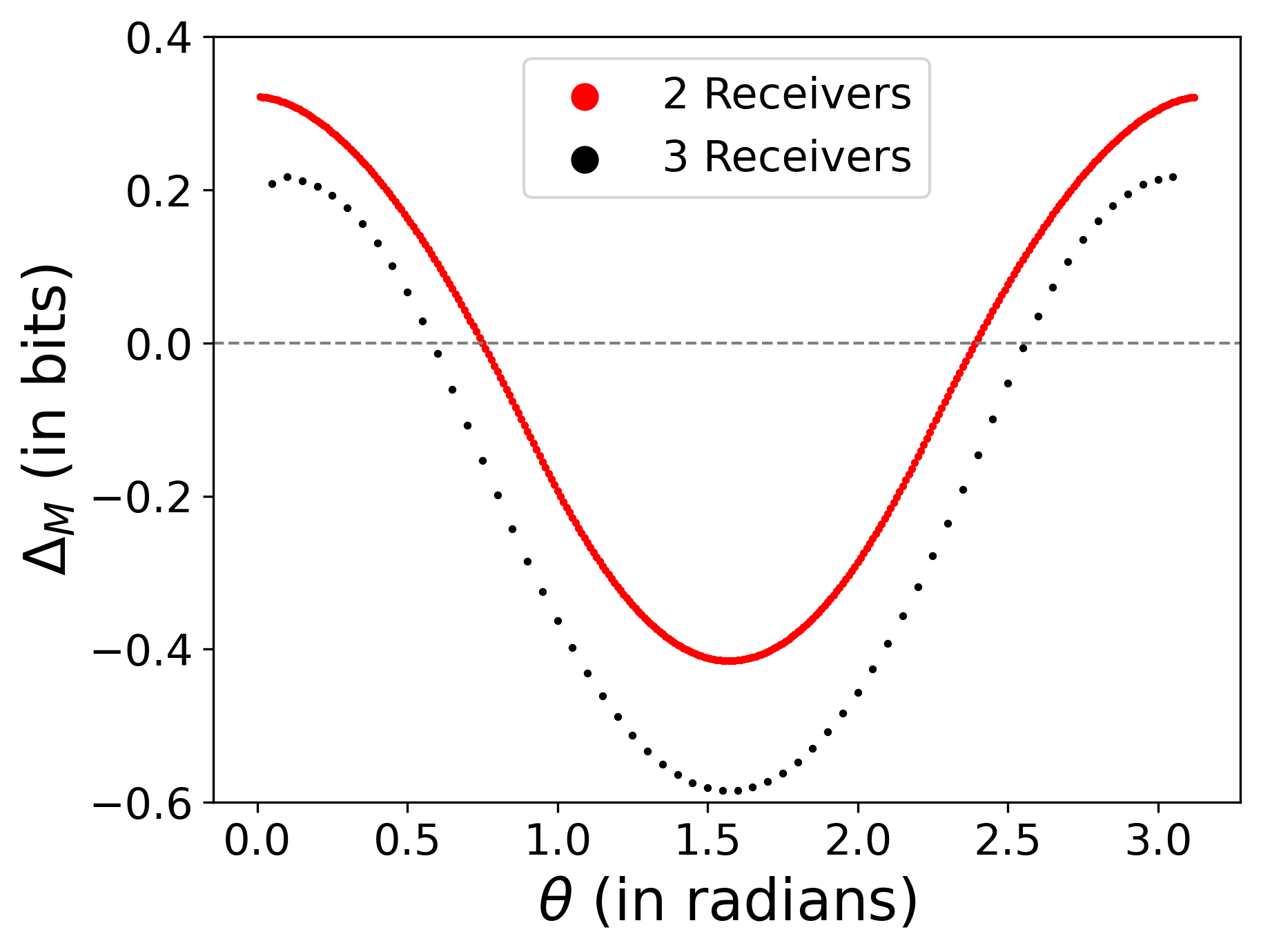}
\caption{The figure of merit chosen for comparing our dense coding protocol with the standard one, given by $\Delta_M \equiv  \chi_{NCR}^{glo} - (\chi_{SDC} + \log{M})$, is plotted along the vertical axis, as a function of $\theta$ of generalised GHZ state, $\ket{gGHZ}_{AB_1\cdots B_M}$, along the horizontal axis. The red solid curve corresponds to $M=2$, and the black dotted curve corresponds to $M=3$. In terms of such a tradeoff, the two-receiver case performs better than the 3-receiver case.} 
\label{fig:GHZ_adv_3receivers}
\end{figure}

Assuming the state shared by the sender and receivers to be $\ket{gGHZ}_{AB_1\cdots B_M}$, a numerical optimisation over all allowed encodings is carried out. In an optimal setting, we find that there exists an encoding and a decoding scheme where, in the asymptotic limit, we have
\begin{equation}
    |\mathcal{X}| \le 2M+2\Rightarrow \widetilde{\chi}_{NCR}^{glo} = \log|\mathcal{X}|~\text{bits}, \;\;\; \text{while}
\end{equation}
\begin{equation}
    |\mathcal{X}| \ge 2M+2\Rightarrow \chi_{NCR}^{glo} = \log(2M+2)~\text{bits}.
\end{equation}
Like earlier, we observe maximum DC capacity even with non-maximally entangled generalised GHZ, as seen in Fig.~\ref{fig:GHZ_DC_theta_3receivers}. Since the control is a $d$-dimensional quantum system, Alice would need to communicate $\log{d}$ bits of information about the communication outcome. So, in this scenario, the advantage over the standard dense coding protocol with multiple receivers is gauged by the quantity, $\Delta_M \equiv \chi_{NCR}^{glo} - (\chi_{SDC} + \log{M})$.  From Fig.~\ref{fig:GHZ_adv_3receivers}, we see that $M=3$ also provides an advantage but not better than the advantage seen in the $M=2$ case. We suspect that, in this regard, $M=2$ is the optimal distributed scenario.

\subsubsection{Unentangled Pure States}
If the shared state is pure, unentangled in the bipartition of $A:B_1B_2...B_M$, then we observe that there exists an encoding and a decoding scheme where, in the asymptotic limit, we have
\begin{equation}
    |\mathcal{X}| \le M+1 \Rightarrow \widetilde{\chi}_{NCR}^{glo} = \log(|\mathcal{X}|)~\text{bits}, \;\; \text{and}
\end{equation}
\begin{equation}
    |\mathcal{X}| \ge M+1 \Rightarrow \chi_{NCR}^{glo} = \log(M+1)~\text{bits}.
\end{equation}
Therefore, the figure of merit defined for comparing this protocol with the standard one is given by $\Delta_M = \chi_{NCR}^{glo}  -(\log{M} + \chi_{SDC} ) = \log(M+1)-\log{M}-1 = \log(\frac{1}{2}+\frac{1}{2M})$. So we find that in this case, $\Delta_M <0~ \forall ~M\in \mathbb{Z}^+$, therefore, there is no advantage with unentangled shared states. 

Therefore, using a higher dimensional quantum system as a control to simultaneously and coherently encode and route the sender's qubit to multiple receiver labs, we enable the dense coding capacity to be a monotonically increasing function of $M$, that is, $\chi_{max} = 1 + \log(M+1)$ bits (for entangled states). The dense coding capacity is $\chi = \log(M+1)$ bits if the state is not entangled.

\section{Conclusion} \label{sec:ending}
Dense coding has long been recognised as one of the cornerstone protocols in quantum communication, enabling the transmission of classical information at rates exceeding classical limits by leveraging quantum resources. While traditional approaches rely on quantum entanglement as a resource, recent advancements have explored the role of additional resources, such as coherent control and indefinite causal order, in enhancing communication protocols. Motivated by these works, we make use of a coherent quantum system as a control in a dense coding protocol and find that the utilisation of a control quantum system helps provide dense coding capacity enhancement under global and one-way LOCC
decoding strategies. 

We leverage a control quantum system to create a superposition of two dense coding scenarios, resulting in a protocol where Alice’s encoding and her choice of routing her system to different labs are coherently governed by the state of the control system. In the global decoding scenario, we numerically maximize the Holevo quantity, $\chi(\mathcal{E})$, of an encoded ensemble, $\mathcal{E}$, over all possible encodings permitted by our protocol to determine the dense coding capacity. Since our protocol requires a noiseless classical channel to communicate the measurement outcome of the control quantum system to the receivers, we impose a constraint that this classical channel has limited capacity, allowing only one bit of information to be transmitted per quantum channel use. To ensure a fair comparison, we account for this additional classical resource in the standard dense coding scenario. To evaluate the advantage of our protocol over standard dense coding, we define the figure of merit as $\Delta \equiv \chi_{NCR}^{glo} - (\chi_{SDC}+1)$, where $\Delta>0$ indicates that our protocol outperforms the standard approach. For a certain set of weakly entangled GHZ states, a completely separable mixed state, and a maximally mixed state, we find $\Delta>0$, demonstrating that our protocol outperforms the standard protocol.

An important feature of these results is that the observed advantage is not restricted to highly entangled resource states. In conventional dense coding, entanglement is typically regarded as the primary resource responsible for communication enhancement. Our results show that a positive advantage can still be obtained for weakly
entangled generalized GHZ states and even for certain
separable and maximally mixed states. This suggests that the coherent superposition of communication routes can provide a communication enhancement that is not solely determined by the amount of entanglement present in the shared resource state. Consequently, useful dense coding advantages may remain accessible even when the shared resource is significantly degraded by noise or possesses little entanglement.

In the LOCC decoding scenario, we simplify the problem by restricting the decoding measurements to a class of one-way LOCC measurements with rank-1 projective measurements, which we denoted as $\texttt{LOCC}_1$. Within this framework, we numerically optimize the mutual information between the sender’s encoding and the receivers’ decoding over all possible encoding strategies and $\texttt{LOCC}_1$ measurements. Our results show that for a nearly unentangled GHZ state, a specific one-shot $\texttt{LOCC}_1$ decoding strategy in our protocol can outperform the standard dense coding protocol even when the latter employs global decoding strategies, which are considered in the asymptotic limit, under the same nearly unentangled GHZ state.

To consider realistic scenarios, instead of assuming a noiseless quantum channel during routing of the state, we extend the protocol's formalism to noisy regimes where coherent control of CPTP maps is realised by coherently controlling over their Stinespring unitaries. We consider quantum dephasing noise and show that our protocol, under this quantum noise, can outperform even the noiseless standard dense coding protocol for dephasing probabilities $p\le0.03$.

We then extend our protocol from two receivers to $M$ receivers, where Alice utilizes an $M$-dimensional quantum system as a control to superpose $M$ different processes. Focusing on the global decoding scenario, we observe that for a GHZ state, the dense coding capacity in our protocol increases monotonically with $M$. Specifically, we find that $\chi_{NCR}^{glo} = 1+ \log{(M+1)}$ for entangled states and $\chi_{NCR}^{glo} = \log{(M+1)}$ for unentangled states in the bipartition of senders to receivers. To fairly compare our protocol with the standard dense coding protocol, we account for the additional classical communication required by our scheme. Since we employ a noiseless classical channel, the figure of merit used for comparison is defined as $\Delta_M \equiv \chi_{NCR}^{glo} - (\chi_{SDC}+\log{M})$. Our results indicate that for low-entangled GHZ states, our protocol outperforms the standard protocol. However, we also find that the $M=2$ case generally performs better than $M=3$. We attribute this to the fact that while $\chi_{NCR}^{glo}$  increases with $M$, the advantage gained is offset by the growing cost of the additional classical communication required to transmit the measurement outcome to the receivers.

\acknowledgements 

We acknowledge computations performed using the \texttt{trust-constr} method in SciPy~\cite{2020SciPy-NMeth, conn2000trust}. A.B. acknowledges support from the INFOSYS Scholarship for Senior Students at Harish-Chandra Research Institute, India. U.S. acknowledges financial support from the Anusandhan National Research Foundation (ANRF), Government of India, under the Grant No. ANRF/ARG/2025/004617/PS.   

\section*{DATA AVAILABILITY}
The data that support the findings of this study are available from the corresponding author upon reasonable request.

\bibliography{ref}

\end{document}